\documentclass[]{aa}

\usepackage[varg]{txfonts}
\usepackage{graphicx,rotating,natbib,color}
\usepackage{amsmath,amssymb}

\begin{document}

\title{Thermal evolution and sintering of chondritic planetesimals}

\subtitle{IV. Temperature dependence of heat conductivity of asteroids and meteorites}

\author{ 
      Hans-Peter Gail\inst{1} 
 \and Mario Trieloff\inst{2,3}
}

\institute{
Zentrum f\"ur Astronomie, Institut f\"ur Theoretische Astrophysik, 
           Heidelberg Univerity, Albert-Ueberle-Str. 2,
           69120 Heidelberg, Germany, \email{gail@uni-heidelberg.de}
\and
Institut f\"ur Geowissenschaften, Universit\"at Heidelberg, Im Neuenheimer
           Feld 236, 69120 Heidelberg, Germany, \email{Mario.Trieloff@geow.uni-heidelberg.de}
\and
Klaus-Tschira-Labor f\"ur Kosmochemie, Universit\"at Heidelberg, Im Neuenheimer Feld 236, 69120 Heidelberg, Germany
  }

\offprints{\tt gail@uni-heidelberg.de}

\date{Received date ; accepted date}

\abstract
{
Understanding the compaction and differentiation of the planetesimals that formed during the initial phases of our solar system and protoplanets from the Asteroid Belt and the terrestrial planet region of the Solar System requires a reliable modelling of their internal thermal evolution. An important ingredient for this is a detailed knowledge of the heat conductivity, $K$, of the chondritic mixture of materials from which planetesimals are formed. The dependence of $K$ on composition and structure of the material was studied in the previous study of this series. For the second important aspect, the dependence of $K$ on temperature, laboratory investigations on a number of meteorites exist concerning the temperature variation of $K$, but no explanation for the observed variation has been given yet.}
{We evaluate the temperature dependence of the heat conductivity of the solid chondritic material from the properties of its mixture components from a theoretical model. This allows to predict the temperature dependent heat conductivity for the full range of observed meteoritic compositions and also for possible other compositions.
}
{Published results on the temperature dependence of heat conductivity of the mineral components found in chondritic material are fitted to the model of Callaway for heat conductivity in solids by phonons. For the Ni,Fe-alloy published laboratory data are used. The heat conductivity of chondritic material then is calculated by means of mixing-rules. The role of micro-cracks is studied which increase the importance of wall-scattering for phonon-based heat conductivity. 
}
{Our model is applied to published data on heat conductivity of individual chondrites. The general trends for the dependency of $K$ on temperature  found in laboratory experiments can largely be reproduced for the set of meteorites if the heat conductivity is calculated for a given composition from the properties of its constituents. It is found that micro-cracks have a significant impact on the temperature dependence of $K$ because of their reduction of phonon scattering length.
}
{}


\maketitle



\section{%
Introduction}

The parent bodies of ordinary chondrites are considered to be undifferentiated planetesimals with diameters estimated to range between about hundred and at most a few hundred kilometres \citep[see, e.g., ][ for a review]{Gai14}. The thermal evolution of such bodies is characterized by a brief initial heating period due to decay of short-lived radioactives (mainly $^{26\!}$Al), lasting for a period of a few million years, followed by a long lasting cooling period of at least 100\,Ma duration, where the heat liberated within the body is transported to its surface where it is radiated away \citep[see, e.g., ][ for a review]{McS03}. It is possible to recover information on these processes by modelling the structure, composition, and thermal history of meteorite parent bodies and by comparing such models with results of laboratory investigations on the composition and  structure of meteorites, and on their thermal evolution as inferred from ages of radioactive decay systems that determine the time when a rock cooled through a characteristic temperature, the so called closure temperature.

A realistic modelling of the raise and fall of temperature during the evolution of the parent bodies requires \emph{inter alia} to use an as accurate as possible value of the heat conduction coefficient, $K$, for solving the heat conduction equation. The present work is the continuation of our attempts to improve the model construction for the thermal evolution of planetesimals \citep{Hen11,Hen12, Hen13, Gai14b, Hen16}. In the last publication we studied the dependence of heat conductivity of chondritic material on mineral composition and porosity. Here we aim to determine the temperature dependence of heat conductivity, $K$, from the properties of the constituents of chondritic material. 

This task is complicated by the rather complex structure of the chondrite material. It is a mixture of many different minerals and a Ni,Fe--alloy with quite different individual properties, and it has a porous granular structure consisting of a two-component mixture from glassy beads of 0.1 to 1\,mm size (the chondrules) and a very fine-grained matrix material with particle sizes of the order of micrometers, also consisting of a variety of materials. This granular material is found in its most pristine form in meteorites of the petrologic class~3 and is assumed to represent the initial structure of the material from which the parent bodies formed. Textural variations defining various petrologic types can be found among all different chemical classes ascribed to different asteroidal parent bodies that can be distinguished by abundance and oxidation state of metal: while H chondrites have a high iron content, L chondrites are low in iron, LL chondrites both low in total Fe and metallic Fe. EH and EL groups are high and low iron clans of enstatite chondrites, which have a rather reduced mineralogy, with no Fe in silicates, and only Fe in sulfides and metal. In higher petrologic classes the initial pore space is partially or completely annealed by sintering. Instead the material has accumulated over time a significant volume fraction of micro-cracks as a result of continued impacts on the surface of the body and the resulting shock waves propagating into the body. Most of the pore-space observed in ordinary chondrites is due to such micro-cracks \citep{Con08}. We found in \citet{Hen16} that the micro-cracks seem to dominate the variation of the measured heat conductivity of the  chondritic material with porosity and to be responsible for its large scatter. Cracks act as obstacles for the propagation of phonons in the lattice of solids and hamper conduction of heat. Also the strong anisotropy of the heat conductivity found in \citet{Ope12} is likely due to a preferred orientation of the cracks formed by passage of a shock wave \citep[e.g.][]{Kov12}.

The heat conductivity of the material is determined essentially by three different processes. First, at temperatures well below 1\,000\,K heat is transported through solids by phonons. Second, starting at temperatures somewhat below 1\,000\,K and increasing rapidly in importance with raising temperature heat is transported by radiative transfer. Third, if in porous material the voids are filled with gas or a fluid material, heat is conducted across the void space by thermal agitation of atoms or molecules. All three processes act independently of each other and require a different kind of treatment. In the present work we concentrate on the heat conductivity by phonons. This is the dominant mode of heat transport up to temperatures of about 700~\dots\ 800\,K where radiative transfer starts to contribute. The pores in ordinary chondrites are generally believed to be empty because parent bodies of chondrites have no atmospheres and there is no contribution to conduction by pore-filling material. The deviating conditions encountered in parent bodies of carbonaceous chondrites (pore filling by ice and/or water during the early evolutionary phase, by swollen hydrated silicates during later evolution) will not be considered in this work.

Experimental data for the temperature variation of heat conductivity of meteorites are available from the work of \citet{Yom83}, \citet{Ope10,Ope12}, and \citet{McC15}. The first one presents results for H and L chondrites for the temperature range between 100 K and 500 K and give convenient fit-formula for the average heat conductivity of chondritic material which have been used in a number of model calculations \citep{Ben96, Akr98,Har10}. The second and third work studies the low-temperature conductivity between a few degrees of Kelvin and room temperature for a variety of meteorite classes.

\begin{figure}

\includegraphics[width=\hsize]{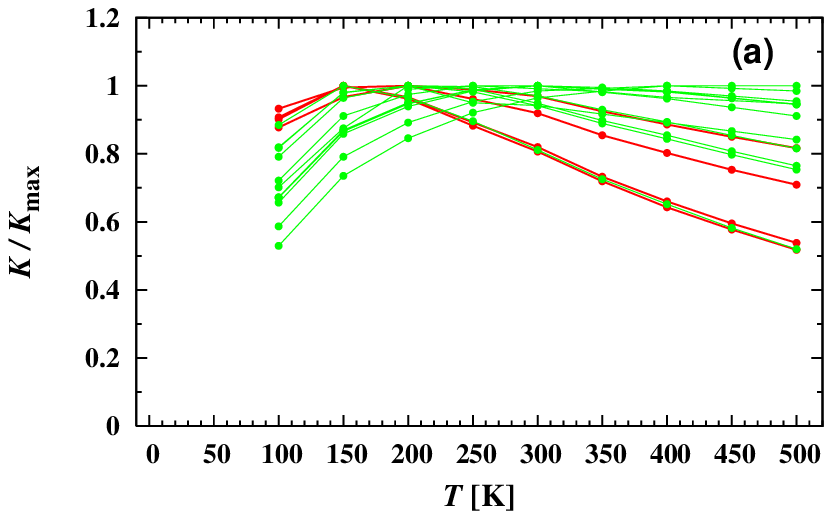}

\includegraphics[width=\hsize]{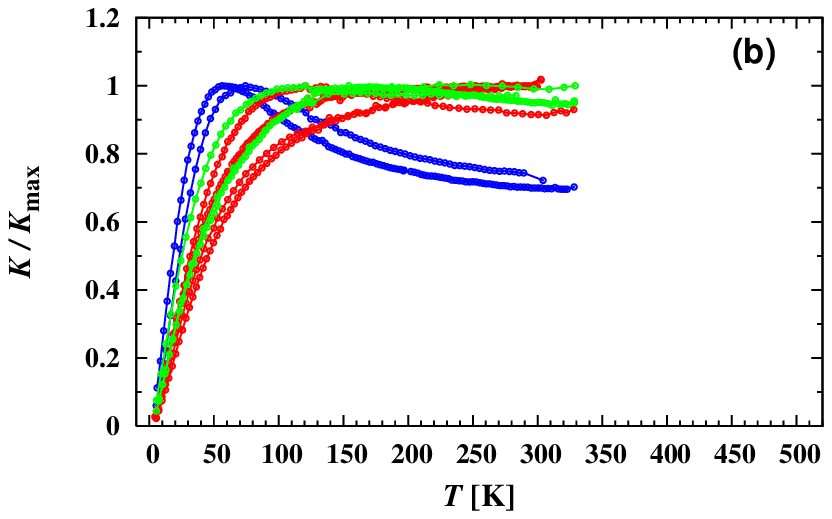}

\includegraphics[width=\hsize]{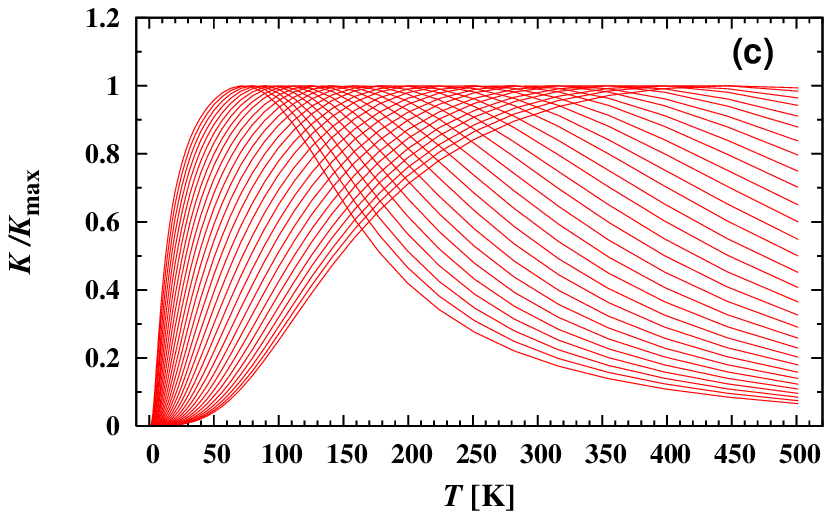}

\medskip

\caption{Temperature variation of heat conductivity of chondritic matter, normalised by its maximum value. (a) Data from \citet{Yom83} for H (red lines) and L (green lines) chondrites. (b) Data from \citet{Ope12} for H (red lines), L (green lines), and E (blue lines) chondrites (data made kindly available to us by G. Consolmagno). (c) Normalized heat conductivity calculated from theoretical model for varying contributions of scattering at walls to phonon scattering, from being negligible (the curves with a pronounced peak at low temperatures) to dominating wall-scattering (curves with flat maximum).
}
 
\label{FigKTvar}
\end{figure}

In the low temperature regime the variation of the heat conductivity with temperature shows in a few cases the strongly peaked nature with a rapid fall-off to zero conductivity in the low temperature range and a less pronounced decrease in the high temperature range, which is well known for pure solids and predicted by the theory of Debye. Most meteorites show, however, an only slow variation of heat conductivity with a broad and weakly pronounced maximum at a few 100\,K (Fig.~\ref{FigKTvar}). This is quite unusual for crystalline materials. 

The note by \citet{McC15} presents some results from 300 K up to over 1\,000\,K. Their data show at the high temperature end already the increasing contribution of radiative transfer. 

We aim to discuss in this work the cause of this different and to some extent unexpected behaviour of $K(T)$ found for most meteorites. We will show that this is a result of the high density of micro-cracks existing in the studied chondrites. The abundant cracks result from the numerous high-velocity impacts on the surface of the meteorite parent body or a major fragment of it and from the last, the excavating, impact. According to theoretical considerations on the gradual growth of regolith layers \citep{Hou79,War02} one expects an only very limited covering of the surface by high-velocity impact craters during the essential initial period of at most 200 Ma for heating and cooling of the parent body. The pristine material from which the parent body originally formed, thus, must have had different heat conduction properties than the chondrite material investigated in the laboratory because of a low density of cracks. Such material seems to be rare in meteorite collections, or at least among the samples presently studied for heat conductivity. We therefore attempt to determine the temperature variation of heat conductivity of the true pristine material from the properties of its constituents and their individual mass fractions.
 

\section{Data for chondrites}

Experimental data for the temperature variation of the heat conductivity of ordinary chondrites are available from two sources. \citet{Yom83} published data for 22 H and L chondrites for a temperature range between 100~K and 500 K in steps of 50 K. \citet{Ope10,Ope12} published data for twelve H, L, E, and C chondrites from a few Kelvin to 300 K in steps of a few Kelvin. The dataset of \citet{Ope10,Ope12} was kindly made available to us by G. Consolmagno. 

These data are plotted in Fig.~\ref{FigKTvar}a,b.  Because the  absolute value of the conductivity strongly depends on the degree of porosity of a sample, the data in the plot are normalized  to the maximum value, $K_{\max}$, of the heat conductivity, in order that the temperature variation of $K$ for different meteorites can be compared. This is possible since in the analytic fit of \citet{Yom83} to their empirical data and in the numerical modelling by \citet{Hen16} it turned out that in good approximation the effect of porosity on the heat conductivity can be factored out, such that $K/K_{\max}$ essentially shows the temperature variation of the compact material. 

At first glance it may appear that Fig.~\ref{FigKTvar}a and Fig.~\ref{FigKTvar}b look rather different. If one refrains, however, from the two enstatite chondrites and from the data below 100 K in Fig.~\ref{FigKTvar}b, there is no big difference in the set of curves shown Fig.~\ref{FigKTvar}b compared to Fig.~\ref{FigKTvar}a  if one considers that the number of H and L chondrites shown in Fig.~\ref{FigKTvar}a is bigger than in Fig.~\ref{FigKTvar}b such that Fig.~\ref{FigKTvar}a represents a broader range of possible properties of H and L chondrites.

The temperature dependences of the heat conductivity show significant variations between different meteoritic specimens, but a general trend seems obvious: (1) There is a broad peak at some temperature $T_\mathrm{p}$. (2) The heat conductivity drops to zero for $T\to0$. (3) The heat conductivity drops much more slowly for increasing temperature at $T>T_\mathrm{p}$. (4) The width of the peak strongly increases with increasing $T_\mathrm{p}$. The broadening and shifting of the temperature maximum shows that besides the temperature-independent reduction of heat conductivity by porosity an additional mechanism is responsible for the modification of the temperature variation of heat conductivity. The similarity of the curves suggests that the different observed temperature-variation profiles are essentially members of a one-parameter family of temperature-variation laws.


\section{Heat conductivity of insulators}
\label{SectDebye}

In this work we discuss which physical parameters determine the observed behaviour of $K(T)$ of chondrites. Their material is dominated by insulating minerals (mainly silicates) and in zero order  approximation the heat conductivity is dominated by such materials. In a first step we therefore neglect the presence of metallic iron, which will be included later.   

\subsection{Model of Callaway}

The phonon thermal conductivity in insulators can be described by the Debye model, where scattering of phonons by surfaces, impurities, and by phonon-phonon interaction determines the propagation of phonons and the resulting transport of heat energy. This theory and the corresponding theory of heat capacity of minerals are discussed in any textbook on solid state physics and will not be discussed.

A version  of the Debye model for heat conduction which has found wide applications was developed by \citet{Cal59}. In this theory the heat conductivity, $K$, can be expressed as a sum of two terms, where usually only one of them is important. The dominant term is given by
\begin{equation}
K={1\over3}cT^3\int_0^{\Theta_{\rm a}/T}{x^4{\rm e}^x\over\left({\rm e}^x-1\right)^2}\,\tau(x,T)\,{\rm d}x
\label{ModelDebye}
\end{equation}
(for the small correction terms not considered here see \citet{Cal59}). Here $T$ is temperature, $\Theta_{\rm a}$ is the Debye temperature of the acoustical branch which characterizes the energy of the phonon energy levels, and $c$ is a constant determined by lattice properties. The integration variable is $x=\hbar\omega/kT$ where omega is the wavenumber of the phonons. We consider only this dominant term. The relaxation time $\tau$ for phonon scattering takes the form
\begin{equation}
\tau=\left({v\over L}+A x^4T^4+Bx^2T^5{\rm e}^{-\Theta_{\rm a}/T}\right)^{-1}
\label{TauScatt}
\end{equation}
with two constants $A$, $B$ given by lattice properties, $v$ being the sound velocity and $L$ a length scale for scattering at boundaries \citep[c.f.][]{Mor06}. The term $\propto A$ describes scattering at impurities and the term $\propto B$ scattering by umklapp processes. The details of such processes are described in any textbook on solid state physics \citep[e.g.][]{Dek65}. The Debye temperature $\Theta_{\rm a}$ is defined by the acoustic branch of the phonon dispersion relation and has to be calculated from this. It can be approximated by
\begin{equation}
\Theta_{\rm a}={\Theta_\mathrm{D}\over n^{1\over3}}\,,
\label{RelDebTemps}
\end{equation}
where $\Theta_\mathrm{D}$ is the Debye temperature entering the Debye theory of the specific heat of solids, and $n$ is the number of atoms per unit cell. Since the specific heat also includes contributions of phonons from the optical branch of the phonon dispersion relation $\Theta_\mathrm{D}$ one always  has $\Theta_{\rm a}< \Theta_{\rm D}$.

\subsection{Wall scattering}

Figure \ref{FigKTvar}c shows as example for fixed $\Theta_\mathrm{a}$ and fixed $A/B$ the variation of $K$ with varying ratio $v/LB$, i.e., with varying importance of wall scattering compared to scattering by umklapp processes. If scattering at walls is unimportant  relative to other phonon scattering processes, the heat conductivity has a pronounced maximum at low temperature and then decreases rapidly with increasing temperature. If scattering at walls is important for scattering of phonons, the heat conductivity has a broad maximum at a much higher temperature and falls off much more slowly toward higher temperatures. In case of meteoritic material one has to observe that the large number of micro-cracks responsible for most of the observed porosity of ordinary chondrites \citep{Con08} acts for phonons as internal walls. For such a material one expects wall scattering to be very important. An inspection of Fig.~\ref{FigKTvar} shows that, in fact, such behaviour is observed in many cases.

The presence of large numbers of micro-cracks and their significant impact on heat conductivity was already inferred by \citet{Yom83}. Their data imply that the micro-cracks are highly flattened and in most cases are randomly oriented and in that case represent a significant obstacle for the heat flux through the material, as is also discussed in \citet{Hen16}. As we will see later, this mode of lowering the heat conductivity acts in a different way with regard to the temperature variation and has to be considered as a separate effect. 


\begin{table*}
\caption{The pure mineral species considered for calculating the properties of
chondrite material, their chemical composition, atomic weight, $A$, mass-density, $\varrho$, thermal conductivity, $K$, at 300 K, Debye temperature $\Theta_\mathrm{D}$ from heat capacity, number of atoms, $n$, per unit cell, estimated Debye temperature, $\Theta_\mathrm{a}$, for heat conductivity according to Eq.~(\ref{RelDebTemps}). The minerals form the indicated solid solutions.}

\begin{tabular}{l@{\hspace{1.0cm}}lrlrlll}
\hline
\noalign{\smallskip}
Solid solution & Chemical & \multicolumn{1}{c}{$A$} & 
   \multicolumn{1}{c}{$\varrho$\tablefootmark{i}} & \multicolumn{1}{c}{$K$} & $\Theta_{\rm D}$\tablefootmark{e} & $n$\tablefootmark{k} & $\Theta_\mathrm{a}$  \\
\quad component & formula & & g/cm$^3$ & W/mK & K &  & K \\
\noalign{\smallskip}
\hline
\\
Quartz             & SiO$_2$             &  60.09 &  2.65   &  6.49\tablefootmark{b}       & 567 &  9  & 273   \\
Olivine            &                     &        &      & & & &   \\
\quad forsterite   & Mg$_2$SiO$_4$       & 140.70 & 3.227 & 5.158\tablefootmark{a}  & 747 & 28 & 246  \\
\quad fayalite     & Fe$_2$SiO$_4$       & 203.77 & 4.402 & 3.161\tablefootmark{a}  & 532\tablefootmark{f} & 28 & 203  \\
Orthopyroxene      &                     &        &      & & & & \\
\quad enstatite    & Mg$_2$Si$_2$O$_6$           & 200.79 & 3.204 & 4.909\tablefootmark{d}  & 719 & 40 & 210  \\
\quad ferrosilite  & Fe$_2$Si$_2$O$_6$           & 263.86 & 4.002 & 3.352\tablefootmark{d}  & 541\tablefootmark{h} & 40 & 158 \\
Clinopyroxene      &                     &        &      & & & & \\
\quad diopside & CaMgSi$_2$O$_6$   & 216.56 & 3.279 & 5.376\tablefootmark{a}  & 654 & 40 & 191  \\
\quad (+ en + fs) \\
Plagioclase        &                     &        &      & & & & \\
\quad albite       & NaAlSi$_3$O$_8$     & 262.23 & 2.610 & 2.349\tablefootmark{a}  & 462 & 52 & 124  \\
\quad anorthite    & CaAl$_2$Si$_2$O$_8$ & 278.36 & 2.765 & 1.679\tablefootmark{a}  & 518 & 104 & 131  \\
\quad orthoclase   & KAlSi$_3$O$_8$      & 278.33 & 2.571 & 2.315\tablefootmark{a}  & & 52 & \\
Troilite           & Fe$_{1-\delta}$S                 &  87.91 & 4.91 & 4.60\tablefootmark{b} & 270\tablefootmark{g} & 1.875 & \\
Iron-nickel alloy  &                     &        &      & & & & \\
\quad iron         & Fe                  &  55.45 & 7.81 & 71.100\tablefootmark{c} & & & \\
\quad nickel       & Ni                  &  58.69 & 8.91 & 80.800\tablefootmark{c} & & & \\
\noalign{\smallskip}
\hline
\end{tabular}

\tablefoot{
\tablefoottext{a}{\citet{Hor69}}, \tablefoottext{b}{\citet{Cla95},} \tablefoottext{c}{\citet{Ho78},} \tablefoottext{d}{extrapolated value from \citet{Hor72},} \tablefoottext{e}{from \citet{Kie80}}, \tablefoottext{g}{from \citet{Chu71}}, \tablefoottext{f}{from \citet{Kob99}}, \tablefoottext{h}{from \citet{Cem06}}, \tablefoottext{i}{from \citet{Smy95}}, \tablefoottext{k}{number of formula units per primitive cell taken from \citet{Smy95}}.
}

\label{TabMinProp}
\end{table*}

\section{%
Heat conductivity of components}

\label{SectKComp}

\subsection{Laboratory data}

An extended body of information on thermal conductivity for minerals of geophysical interest is available.  The laboratory determined data have been reviewed by, e.g., \citet{Cla95}. For the heat conductivity of the nickel iron alloy one has a very detailed tabulation in \cite{Ho78} for the temperature range from 4\,K to 1\,100\,K and for a large set of mole fractions of Ni in the alloy. 

For the mineral components one encounters the problem, that virtually no data on heat conductivity for the relevant minerals seem to be available in the temperature range below room temperature. For applications in geophysics essentially only temperatures $T\gtrsim300$\,K are of interest and a large body of such data for all kinds of materials is available. However, because of their location in the asteroid belt, the parent bodies of meteorites are expected to have surface temperatures significantly below room temperature. All models for their thermal evolution matching the meteoritic record require surface temperatures lower than $300$\,K \citep[cf.][ for a review]{Gai14}. The laboratory data on $K(T)$ at low temperature available for some meteorites \citep{Ope10,Ope12} indicate that there is a significant temperature variation of the heat conductivity below 300\,K. In order to account for the variation in this temperature region we take recourse to a semi-empirical approach based on the classical Debye model for heat conductivity by phonons discussed in Sect.~\ref{SectDebye}. It would be extremely useful if the heat conductivity of the main mineral components of chondrites would be measured in the low-temperature range as in \citet{Ope12}.

In the temperature range above room temperature the heat conductivity of rock minerals generally decreases with increasing temperature \citep[e.g.][]{Rob88,Xu04}, as it is expected if heat conduction is due to phonons (see any textbook on solid state physics). However, for high temperatures, typically from about 800\,K \dots\ 1\,000\,K on, published data on laboratory measured heat conductivities of most minerals frequently tend to increase again due to an increasing contribution of radiative transfer to heat conduction (cf., e.g., Fig. \ref{FigFitOli}a). Since we aim to include heat conduction by radiative transfer in our model calculations as a separate process, laboratory measured data can only be used from the temperature regime where radiation does not contribute to heat conduction. Therefore we use experimental data only for temperatures below 600\,K where phonon conduction dominates. Our procedure then is to fit the Debye model, Eq.~(\ref{ModelDebye}), to the available empirical data in the temperature range below about 600\,K and extrapolate the phonon heat conductivity to higher temperatures by the Debye model.

\subsection{Model fit for $K(T)$}

The Debye model for $K(T)$ depends on five constants: $\Theta_\mathrm{a}$, $c$, $A$, $B$, and $v/L$. In principle, they are determined by the properties of the solid as described in \citet{Cal59}. In practice they cannot be calculated directly from lattice properties except for special cases where all the required input data are available. Therefore we consider the quantities $\Theta_\mathrm{a}$, $c$, $A$, $B$, and $v/L$ as free parameters. and determine them from fitting expression (\ref{ModelDebye}) to a set of measured values of $K(T)$ at different temperatures by the method of least squares. Unfortunately the Debye temperature $\Theta_\mathrm{a}$ determines the temperature variation of $K(T)$ only at low temperatures where no data for the materials of interest could be found. Hence $\Theta_{\rm a}$ presently cannot reliably be determined by a fit to the available measured heat conductivity data. In addition we estimate $\Theta_{\rm a}$ from the Debye temperature $\Theta$ of the heat capacity via Eq.~(\ref{RelDebTemps}), 

\begin{table*}[t]
\caption{Coefficients of the Callaway-model of heat conductivity (resulting $K$ in units W/mK) and the value of $\chi^2$ for the optimum fit. }

\begin{tabular}{lcccc}
\hline\hline
\noalign{\smallskip}
Mineral		& $p_1$ & $p_2$ & $K_0$ & $\chi^2$ \\
\noalign{\smallskip}
\hline
\noalign{\smallskip}
Forsterite  & $2.9622\times 10^{-3}$ & $3.9559\times 10^1$ & $2.5618\times 10^1$ & $7.02\times 10^{-5}$ \\
Enstatite   & $7.6699\times 10^{-4}$ & $1.7397\times 10^1$ & $1.8014\times 10^{1}$ & $8.70\times 10^{-5}$ \\
Diopside    & $7.1231\times 10^{-4}$ & $1.7411\times 10^1$ & $2.1460\times 10^1$    & $9.48\times 10^{-5}$ \\ 
Albite      & $1.1197\times 10^{-3}$ & $1.7071\times 10^3$ & $1.5737\times 10^2$    & $1.66\times 10^{-5}$ \\
\noalign{\smallskip}
\hline
\end{tabular}
\label{TabVarParm1}
\end{table*}

For practical purposes we write Eq.~(\ref{ModelDebye}) as
\begin{align}
K(T)=&K_0\left(T^3\over\Theta_{\rm a}^3\right)\int_0^{\Theta_{\rm a}/T} {x^4{\rm e}^x\over\left({\rm e}^x-1\right)^2}\nonumber\\
&\times\left[p_1+p_2\left(T\over\Theta_{\rm a}\right)^4x^4+\left(T\over\Theta_{\rm a}\right)^3{\rm e}^{-\Theta_{\rm a}/T}x^2\ \right]^{-1}\,{\rm d}x
\label{ModelDebyeFit}
\end{align}
where we have condensed the constants in Eq.~(\ref{ModelDebye}) in 
three new parameters $K_0$, $p_1$, $p_2$ that have to be determined by a fit to the data. The integral is calculated by numerical integration. The fit is done by searching for the smallest value of
\begin{equation}
\chi^2=\sum_ig_i\left(K_i-K(T_i)\right)^2\,,
\end{equation}
where $K_i$ are given values of the heat conductivity at temperatures $T_i$ and $g_i$ are weight factors. The minimum-search is done by a genetic algorithm in the special form described in \citet{Cha95}.

\subsection{The individual minerals}

The main solids found in ordinary chondrites are listed in Table~\ref{TabMinProp} together with some basic data. The table also shows the corresponding Debye temperature $\Theta_\mathrm{a}$ used in the fit procedure for all minerals. Chondrites contain additionally many other mineral components of low abundance which are probably unimportant for the heat conductivity of the mixture; they are neglected.

\subsubsection{Olivine}

Olivine is a solid solution of the end members forsterite and fayalite. Published data  for $K(T)$ for olivine with different compositions $x$ (= mole fractions) are discussed, e.g., in \citet{Rob88} and \citet{Xu04}. The variation of $K$ with composition $x$ is shown for three different temperatures in  \citet{Rob88} and for room temperature in \citet{Hor72}. Since the kind of compositional variation at fixed temperature seems to depend only slightly on temperature \citep{Rob88} we assume that the compositional variation for all temperatures can be factored out such that we can write
\begin{equation}
K(T,x)=K(T,x_0)\cdot k(x)\,,
\label{DefKthXOl}
\end{equation}
where $K(T,x_0)$ is the heat conductivity for some fixed mixture $x_0$, and $k(x)$ describes the variation relative to the reference composition $x_0$. We determine $k$ from the data at room temperature given in \citet{Hor72}. Then in principle we need to determine $K(T,x)$ for only one mixing ratio.

For forsterite and enstatite \citet{Hof99} has developed a model for the phonon heat conductivity at temperatures from 300\,K upward which reproduces empirical data. In this approach, infrared reflectivity/absorptivity data for the minerals are used to determine the density of phonon-states and to calculate from this the high-temperature phonon heat conductivity from heat conductivity at room temperature. We fit a Debye-model to this model in the temperature range $T\ge300$\,K.

The Debye temperatures of forsterite and fayalite for heat conduction, $\Theta_{\rm a}$, is estimated from the Debye temperature of the Debye model for the heat capacity, $\Theta_\mathrm{D}$, by means of relation (\ref{RelDebTemps}). The Debye temperature $\Theta_\mathrm{D}$ is taken from the compilation by \citet{Kie80} and from \citet{Chu71} who gives the variation of $\Theta_\mathrm{D}$ with composition of the solid solution. The number of atoms per unit cell, $n$, in Eq.~(\ref{RelDebTemps}) is determined from the number of atoms of the formula unit (7 for olivine) and from the number of such units forming the primitive cell of the crystal \citep[4 for olivine, see, e.g.,][]{Smy95}.

\begin{figure*}[t]

\includegraphics[width=.49\hsize]{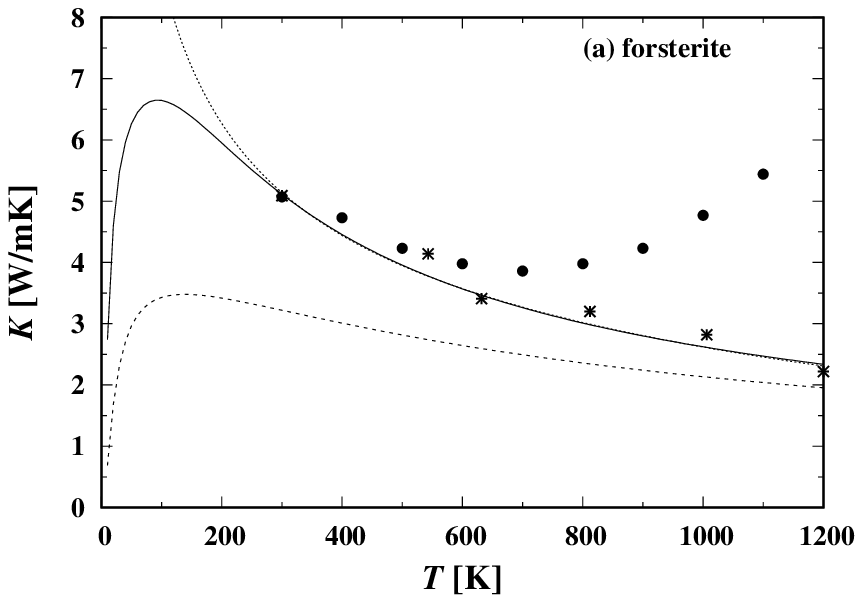}
\includegraphics[width=.49\hsize]{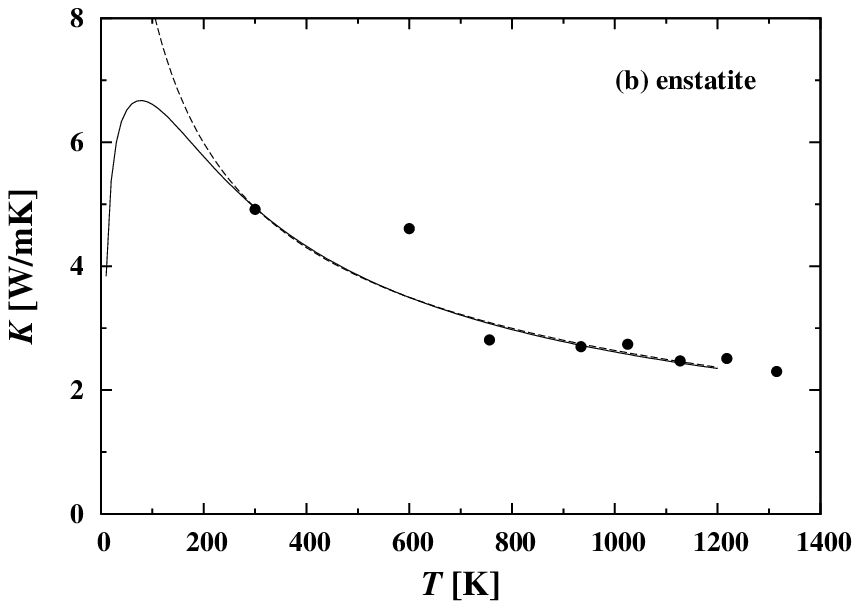}

\includegraphics[width=.49\hsize]{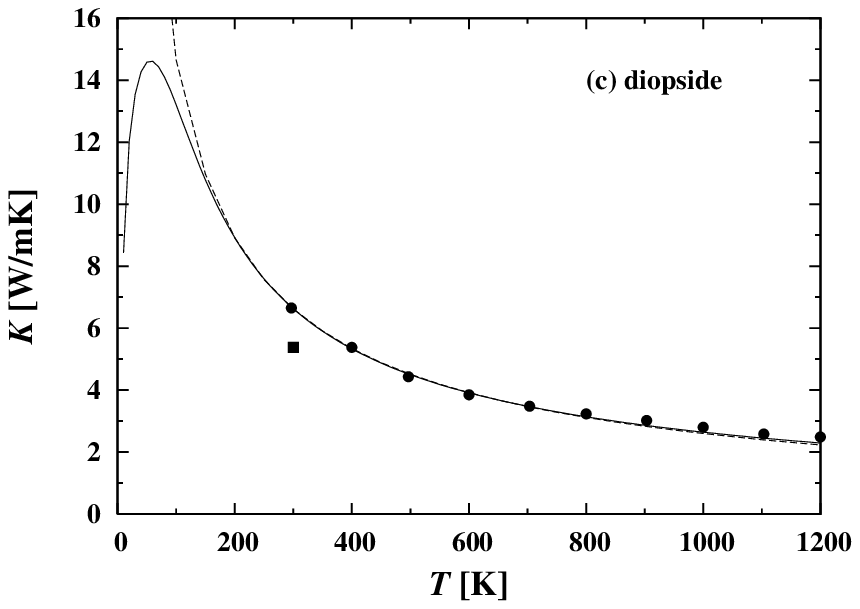}
\includegraphics[width=.49\hsize]{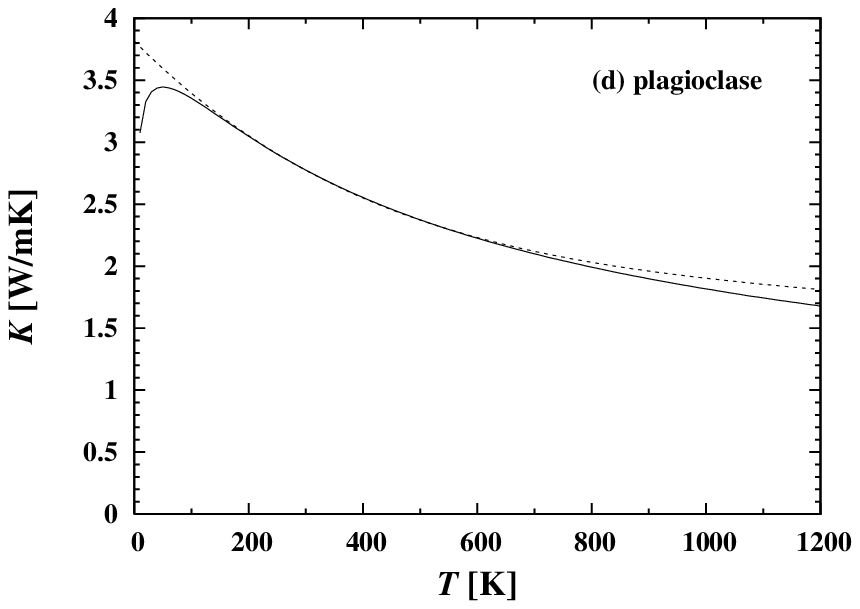}

\caption{Heat conductivity of the mineral components by phonons. \emph{Solid line:} The Callaway model fitted as described in the text. \emph{Dashed line:} \citet{Hof99} model for high-temperature conductivity. (a) Forsterite. \emph{Black squares:} Experimental data for total heat conductivity (including radiative contributions) of olivine with composition Fa$_{\rm 18}$ \citep[from the compilation of][]{Cla95}. \emph{Crosses:} Data for phonon heat conductivity of forsterite from \citet{Sch72}, datum for 300 K from \citet{Cha96}. The dashed line shows the reduced heat conductivity for particles of about 1\,mm diameter. (b) Enstatite. \emph{Black squares:} Experimental data for phonon heat conductivity of orthopyroxene from \citet{Sch72}, datum for 300 K from \citet{Hor72}. (c) Diopside. \emph{Black circles:} Experimental data for phonon heat conductivity of orthopyroxene from \citet{Hof07}.  \emph{Black square:} Datum for 300 K from \citet{Hor69}. (d) Plagioclase. \emph{Dashed line:} Fit to experimental data for heat conductivity of plagioclase given by \citet{Hof09} for the temperature rage between 300\,K and 1\,200\,K.
}

\label{FigFitOli}
\label{FigFitEns}
\label{FigFitDio}
\label{FigFitPla}
\end{figure*}

The fit was done by calculating for temperatures from 300 to 750 K in steps of 25\,K the heat conductivity of forsterite in the approach of \citet{Hof99} and fitting the approximation (\ref{ModelDebyeFit}) to these data by a least square method by searching for the minimum by an evolution algorithm \citep[in the implementation given in][]{Cha95}. The result is shown in Fig.~\ref{FigFitOli}a and the resulting optimum parameters are given in Table~\ref{TabVarParm1}. The model is also compared to experimental data. One set of data (filled circles) show data for the total heat conductivity of natural olivine of composition Fa$_{18}$ which shows the increase to higher temperatures by radiative contributions. The data points at 300, 400, and 500\,K correspond essentially to the conduction by phonons. 

The crosses show the contribution of heat conduction alone as determined by \citet{Sch72} for sintered forsterite. These data correspond to an effective heat conduction coefficient of a polycrystalline material where one has averaged over the anisotropic heat conduction tensor \citep[cf.][]{Cha96} of crystalline forsterite. This is just what one needs for modelling chondritic material.  The data are augmented by a single value at 300 K from \citet{Cha96} which corresponds to the average of the values along  the three main crystal axis.

The data from \citet{Hor72} for the variation of heat conductivity of olivine containing a mole fraction $x$ of fayalite relative to the heat conductivity of pure forsterite can be fitted by \citep[cf.][]{Hen16}
\begin{equation}
k_\mathrm{ol}(x)=1.0000-1.0776\,x+0.67633\,x^2\,,
\end{equation}
which defines the factor $k(x)$ in Eq.~(\ref{DefKthXOl}). This has to be combined with the heat conductivity of forsterite to obtain the heat conductivity of olivine.
 
\subsubsection{Pyroxene}

Pyroxene is a solid solution of mainly enstatite (MgSiO$_3$) and ferrosilite (FeSiO$_3$) with some wollastonite (CaSiO$_3$). In ordinary chondrites it is generally found as two co-existing phases, a Ca-poor phase, the orthopyroxene with enstatite and ferrosilite as the main solution components, and a Ca-rich phase, the clinopyroxene with ferrosilite and wollastonite as the main solution components. Orthopyroxene and clinopyroxene have somewhat different lattice structures. The clinopyroxene is of much lower abundance than the orthopyroxene because of the low Ca abundance compared to Mg in the cosmic element mixture.

\paragraph{Orthopyroxene.}
In the following we consider only enstatite and ferrosilite as solution components in orthopyroxene. We assume that minor other solution components that are inevitably present in natural minerals have no major influence on the heat conductivity of the material. The variation of the heat conductivity of orthopyroxene with $x$, the mole fraction of the iron-rich end-member ferrosilite in the solid solution series, was determined for room temperature by \citet{Hor72}. It can be fitted by \citep[cf.][]{Hen16}
\begin{equation}
K(x)=4.9094-5.0649\,x+3.5078\,x^2\,.
\label{FitKthPyr} 
\end{equation}
The data given for the pure end-members in Table~\ref{TabMinProp} are calculated from this. Other data for room temperature given in \citet{Hor69} and \citet{Cla95} are in accord with the fit (\ref{FitKthPyr}) within the frame of typical accuracies of $\pm10$\% for heat conduction data of minerals as suggested by the data given in the tabulation of \citet{Cla95}.

The dependency on composition given by Eq.~(\ref{FitKthPyr}) is very similar to that in the case of olivine. For other temperatures than room temperature no information on the composition dependency of heat conduction could be found. We assume that also for pyroxene a factorisation like Eq.~(\ref{DefKthXOl}) can be applied. The corresponding function $k_\mathrm{py}(x)$ for the variation with respect to enstatite is given by
\begin{equation}
k_\mathrm{py}(x)=1.0000-1.0317\,x+0.7145\,x^2\,.
\end{equation}
The knowledge of the temperature variation of heat conductivity for one composition then suffices to calculate the heat conductivity for all mixing ratios.

Unfortunately the available data material to determine $K(T)$ for orthopyroxene seems to be scarce. Therefore we follow the same procedure as in the case of forsterite and determine at temperatures $\ge300$\,K the temperature variation of the heat conductivity for the pure end-member enstatite of the orthopyroxene series from the model of \citet{Hof99}. In the work of Hofmeister no explicit value for their parameter $a$ is given for enstatite, but the model was applied by \citet{Gib03} with success to interpret their measurements of high-temperature conductivity of silicate rock minerals where they obtained almost the same value for $a$ for mixtures of pyroxene and olivine as Hofmeister derived it for forsterite. Therefore it seems justified to use for enstatite the same parameter $a$ as obtained by Hofmeister for forsterite. 

The run of heat conductivity of enstatite calculated with these assumptions is shown in Fig.~\ref{FigFitEns}b as dashed line, where also the extrapolation to temperatures $T<300$\,K is shown. Also some laboratory measured data from \citet{Sch72} are shown for comparison, but it is to be observed that the data are for crystalline material and that the material is not pure enstatite. Nevertheless, despite of the deviating point at 600\,K, the general trend for the temperature variation of the Hofmeister-model and the experimental findings appear to concur.

In order to extend the heat conductivity model to low temperatures by means of the Callaway model, the Debye temperature for heat conduction, $\Theta_{\rm a}$, has to be fixed. This is estimated for enstatite and ferrosilite from the Debye temperature of the Debye model for the heat capacity, $\Theta_\mathrm{D}$, by means of relation (\ref{RelDebTemps}). The Debye temperature $\Theta_\mathrm{D}$ is taken for enstatite from the compilation by \citet{Kie80} and for ferrosilite from \citet{Cem06}. The number of atoms per unit cell, $n$, in Eq.~(\ref{RelDebTemps}) is determined from the number of atoms of the formula unit (10 for orthopyroxene) and from the number of such units forming the primitive cell of the crystal \citep[4 for orthopyroxene, see, e.g., ][]{Smy95}.

The Callaway-model is fitted to the high temperature data as in the case of olivine. The resulting variation of heat conductivity with temperature for the whole temperature range $T\ge25$\,K is shown in Fig.~\ref{FigFitEns}b as solid line and the resulting optimum parameters are given in Table~\ref{TabVarParm1}.

\paragraph{Clinopyroxene.}
The composition of the Ca-rich pyroxene in ordinary meteorites varies somewhat between meteorite classes, but since this is an only minor component in the mineral mix we ignore such variations and use for simplicity only a single composition, that of diopside, for all types to calculate the heat conductivity coefficient of ordinary chondrites. The small iron content of the clinopyroxenes will probably result in a somewhat reduced heat conductivity compared to pure diopside, analogous to other silicates \citep[cf.][]{Hor69}, but we neglect also this complication.

The heat conductivity of diopside at room temperature is given in \citet{Hor69}. If one takes the average of the values for the two specimens with a density close to that of pure diopside there results a value of $K=5.38$\,W/mK. Values for the lattice contribution to the heat conductivity of diopside (precise composition was not reported) for the temperature range 300\,K to 1\,900\,K were determined by \citet{Hof07}. The corresponding data are shown in Fig.~\ref{FigFitDio}c. The value at room temperature given by them, $K=6.65$\,W/mK, is somewhat higher, but is likely within of the probably significant error limits of the older measurements. For temperatures $T>300$\,K we approximate the heat conductivity by the model of \citet{Hof99} with a value of their parameter $a$ of 0.7, shown as dashed line in Fig.~\ref{FigFitDio}c. 

For extending the heat conductivity model to low temperatures by means of the Callaway model, the Debye temperature for heat conduction, $\Theta_{\rm a}$, is estimated from the Debye temperature, $\Theta_\mathrm{D}$, of the Debye model for the heat capacity \citep[taken from][]{Kie80} by means of relation (\ref{RelDebTemps}).  The number of atoms per unit cell, $n$, in Eq.~(\ref{RelDebTemps}) is determined from the number of atoms of the formula unit (10 for clinopyroxene) and from the number of such units forming the primitive cell of the crystal \citep[4 for orthopyroxene, see, e.g., ][]{Smy95}. The Callaway-model is fitted to the high temperature data as in the case of olivine. The resulting variation of heat conductivity with temperature for the whole temperature range $T\ge25$\,K is shown in Fig.~\ref{FigFitDio}c as solid line and the resulting optimum fit parameters are given in Table~\ref{TabVarParm1}.

\subsubsection{Plagioclase}

The feldspars in  meteorites are solid solutions of mainly albite with an anorthite mole fraction of 0.12 to 0.15 and a small fraction of orthoclase \citep[cf. Table 3 in][ and references therein]{Hen16}. The thermal conductivity of the main component, albite, was recently determined by \citet{Hof09} and an analytic fit to the experimental data for the temperature region from 300\,K to 1\,200\,K was given. We use in this work this recent new determination of $K(T)$ for albite since older data \citep[reviewed by][]{Rob88} show a deviating behaviour with varying temperature compared to other pure minerals while the data from \citet{Hof09} follow the general trends observed for  minerals. This discrepancy  may result from different compositions and properties of the older investigated samples.   

For obtaining an estimation of heat conductivity of albite at low temperatures by means of the Callaway model, the Debye temperature for heat conduction, $\Theta_{\rm a}$, is estimated from the Debye temperature, $\Theta_\mathrm{D}$, of the Debye model for the heat capacity \citep[taken from][]{Kie80} by means of relation (\ref{RelDebTemps}). The number of atoms per unit cell, $n$, in Eq.~(\ref{RelDebTemps}) is determined from the number of atoms of the formula unit (13 for albite) and from the number of such units forming the primitive cell of the crystal \citep[4 for albite, see, e.g., ][]{Smy95}. The Callaway-model is fitted to the high temperature analytic fit given by \citet{Hof09} (dashed line in Fig.~\ref{FigFitPla}d) similar as in the case of olivine. The resulting variation of heat conductivity with temperature for the whole temperature range $T\ge25$\,K is shown in Fig.~\ref{FigFitPla}d as solid line. The optimum fit parameters for Eq.~(\ref{ModelDebyeFit}) are given in Tab
 le~\ref{TabVarParm1}.

\begin{figure}[t]

\includegraphics[width=\hsize]{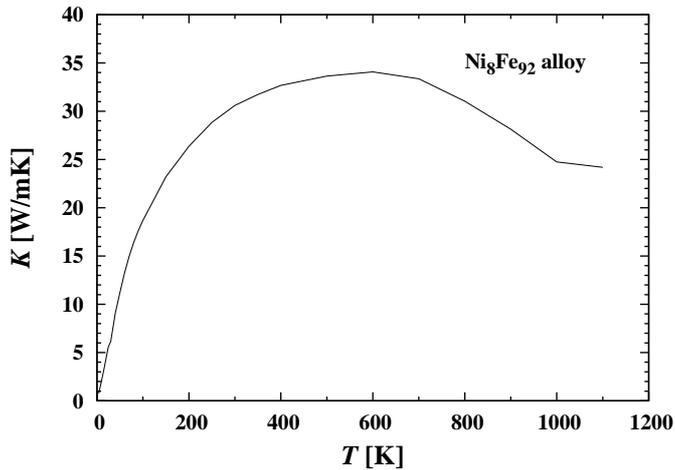}

\bigskip

\caption{Heat conductivity of a nickel-iron alloy with eight mole percent nickel content by phonon and electron conduction \citep[interpolated from the table in][]{Ho78}.}

\label{FigFitNiFe}
\end{figure}

For plagioclase we assume that its heat conductivity is only determined by the two main solution components, albite and anorthite. The influence of minor components is assumed to be negligible. As in the case of olivine we assume that the variation of the heat conductivity with temperature and composition can be described by a factorisation of the kind given by Eq.~(\ref{DefKthXOl}). From the plot given in \citet{Rob88} for the compositional variation of $K(T)$ of feldspars at three different temperatures one infers that this is approximately valid for the anorthite mole fractions of interest ($x\lesssim0.3$) but is not applicable for higher anorthite contents. For room temperature \citet{Hor72} gave data for the variation of heat conductivity of plagioclase containing a mole fraction $x$ of anorthite which can be fitted by a polynomial expression \citep[cf.][]{Hen16} from which one finds
\begin{equation}
k(x)=1-1.32316\,x+1.22475\,x^2\,,
\end{equation}
This has to be combined with the heat conductivity of albite to obtain the heat conductivity of plagioclase with mol fraction $x$ of anorthite.

\subsubsection{Nickel-Iron}

In \citet{Ho78} heat conductivities of Fe-Ni alloys are given for a large variety of Ni contents and temperatures from about 4 K to 1\,000\,K. The appropriate values of $K$ for the nickel content of the iron in  chondrites were obtained by interpolation in this table. Figure \ref{FigFitNiFe} shows the variation of $K(T)$ for a nickel content of eight mole percent, appropriate for H chondrites or enstatite chondrites.

\subsubsection{Troilite}

Only a value for the heat conductivity at room temperature could be found \citep{Cla95}, and data for two temperatures \citep{Tsa82} for the related compound pyrrhotite (Fe$_7$S$_8$). This is insufficient to construct the temperature dependent heat conductivity for this component of the chondritic mixture.


\begin{table}

\caption{Components of the mixture of minerals and metal considered in the modelling of H, L, and EH chondrites, their volume fraction in the mixture, $f$, the mole fraction, $x$, in case of binary solid solutions, and the mass density, $\varrho$, of the mixture.}

\begin{tabular}{@{}lcccccc@{}}
\hline
\hline
\noalign{\smallskip}
 &\multicolumn{2}{c}{H} & \multicolumn{2}{c}{L} & \multicolumn{2}{c}{EH} \\[.1cm]
Component & $f$ & $x$ & $f$ & $x$ & $f$ & $x$ \\
\noalign{\smallskip}
\hline
\noalign{\smallskip}
Olivine       & 0.399 & 0.20 & 0.491 & 0.25 & 0.000 & 0.00 \\
Pyroxene      & 0.291 & 0.17 & 0.253 & 0.22 & 0.666 & 0.00 \\
Clinopyroxene & 0.061 &      & 0.070 &      & 0.000 &       \\
Plagioclase   & 0.114 & 0.18 & 0.109 & 0.16 & 0.144 & 0.19  \\
Nickel-iron   & 0.096 & 0.08 & 0.041 & 0.13 & 0.120 & 0.08  \\
Troilite      & 0.039 &      & 0.036 &      & 0.070 &       \\
\noalign{\smallskip}
Density $\varrho$ & \multicolumn{2}{c}{3.78} & \multicolumn{2}{c}{3.59} & \multicolumn{2}{c}{3.80} \\
\noalign{\smallskip}
\hline
\end{tabular}
\tablefoot{
The mole fractions $x$ refer to the fayalite and ferrosilite in the case of olivine and pyroxene, respectively, to the anorthite component in case of plagioclase, and to the nickel content of the nickel-iron alloy.
\\
Data taken from the compilation in \citet{Hen16}.
}

\label{TabMetComp}
\end{table}

\section{%
Heat conductivity of chondritic mineral mixture}

\subsection{Effective heat conductivity of mineral composites}

Ordinary chondrites of types H, L, LL, EH, and EL are formed by a mixture of minerals and some  fraction of Ni,Fe-alloy in varying proportions for the different types. Data on the average volume fraction of the most abundant mixture components in ordinary chondrites and the average composition of those components that themselves are solid solutions of a number of components are compiled in \citet{Hen16}. In that work the dependency of heat conductivity at room temperature (300 K) on porosity is studied. The same data on volume fraction and average compositions of the components are used here to calculate the variation of heat conductivity with temperature. The essential data used in this work are listed in Table~\ref{TabMetComp}.

Besides the main mixture components listed in Table~\ref{TabMinProp} meteorites contain numerous additional minor components with low abundances. These are neglected because they have no major impact on the average heat conductivity of the mixture. Table 3 in  \citet{Hen16} shows that ordinary chondrites contain a few volume percent of troilite. Unfortunately no sufficient data could be found in the literature on the temperature variation of its heat conductivity. Because its heat conductivity at 300 K is within the range of typical values found for the silicate minerals at this temperature (cf. Table \ref{TabMinProp}) this small mixture component is also neglected, because its inclusion is unlikely to result in noticeable changes of the average conductivity.
In \citet{Hen16} it was found that the effective heat conductivity, $K_\mathrm{eff}$, of chondritic material is rather accurately reproduced by the Bruggeman mixing rule if compared to a direct numerical solution of the heat conduction equation for a test volume filled with a mineral mixture. 
In this approximation one calculates $K_\mathrm{eff}$ from the equation  \citep[cf. the review of ][]{Ber95}
\begin{equation}
K_{\rm eff} = \left( \sum^{N}_{i=1} \frac{3f_i}{2K_{\rm eff} + K_i} \right)^{-1}\,,
\label{EffMediumBrugg}
\end{equation}
where $K_i$ are the heat conductivities of the mixture components, $f_i$ their volume fractions in the mixture, and $N$ is the number of components.  

\begin{figure}

\includegraphics[width=\hsize]{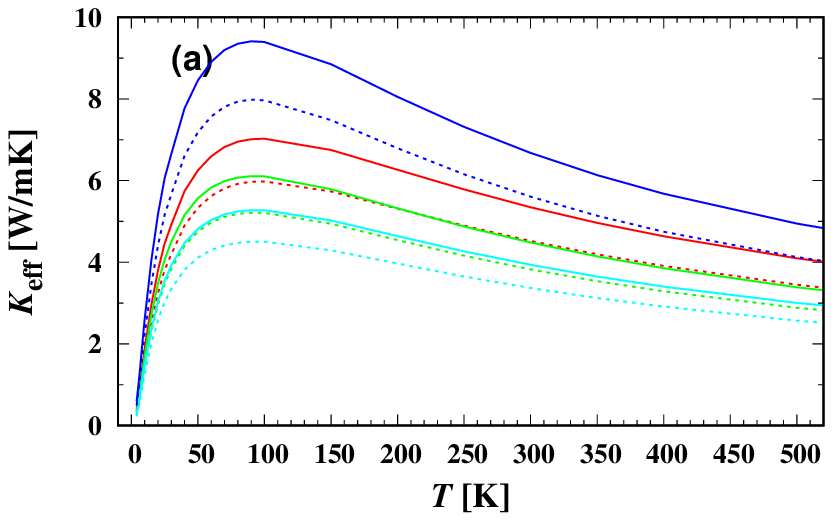}

\includegraphics[width=\hsize]{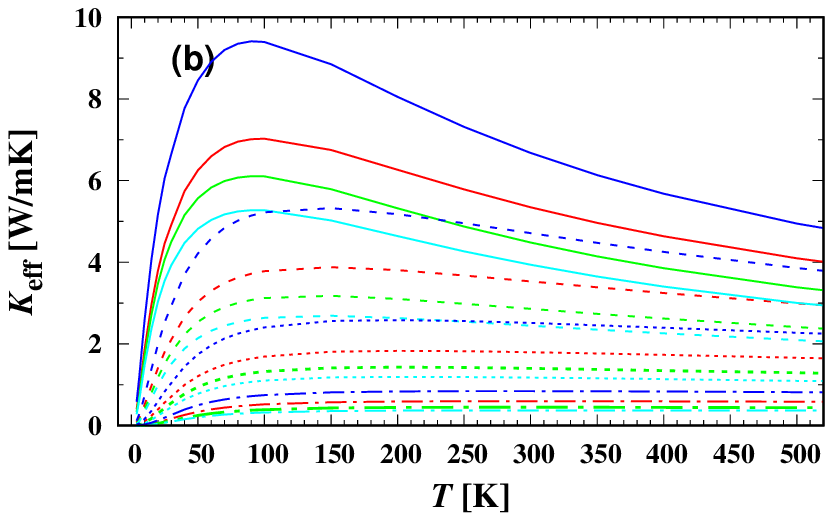}

\includegraphics[width=\hsize]{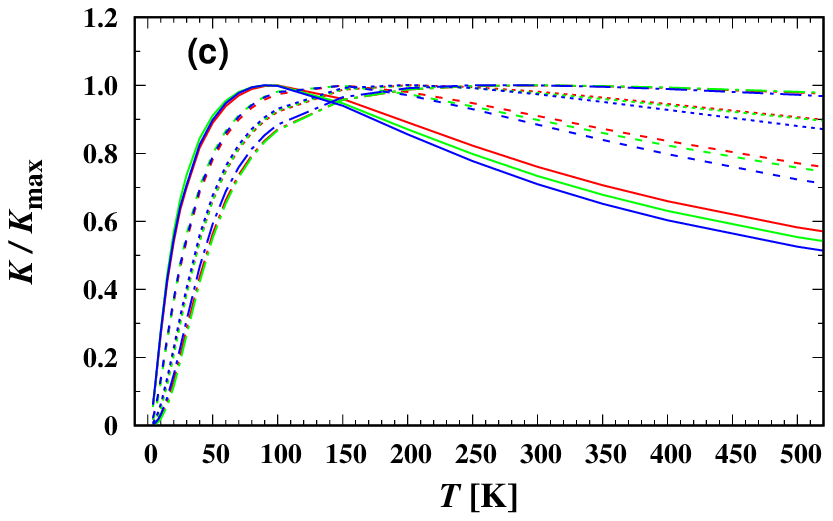}

\caption{Temperature variation of effective heat conductivity of chondritic material for types  H (red), L (green), LL (cyan), and EH (blue). (a) Effect of porosity. The solid line corresponds to fully compacted material, the dashed line to material with 10\% porosity. (b) Variation with enhanced scattering at surfaces or interfaces. Solid lines correspond to the contribution of the scattering term in Eq. (\ref{ModelDebyeFit}) as derived from laboratory data, dashed lines to enhancement of scattering by a factor of 10 (long dashed), factor of 100 (short dashed), and a factor of 1000 (dash-dot). (c) As before, but values of $K_\mathrm{eff}$ normalized by its maximum value. 
}

\label{FigKTcomp}

\end{figure}

We use the Bruggeman mixing rule to calculate the effective heat conductivity of H, L, LL, and EH chondrite material from the temperature-dependent heat conductivities of the mixture components which we discussed in Sect.~\ref{SectKComp}. The result is shown for the temperature range from 10 to 500 K in Fig. \ref{FigKTcomp}a. This range covers the range for which empirical data for some meteorites are available from  \citet{Yom83} and \citet{Ope10,Ope12}. The full lines correspond to completely compacted material without any porosity. In the parent bodies of meteorites such material exists in the inner part of the bodies where temperatures during the thermal history increased to such a value \citep[$T\gtrsim600\dots900$ K, see ][]{Gai14b} that any pore space of the initially granular material is annealed by sintering.

In the layers close to the surface the temperature always remains below the temperature required for complete sintering and part or all of the initial porosity persists in the material. Additionally, in the surface region the material may be subject to shocks from impacts which cause the formation of numerous micro-cracks and lattice defects in the crystal structure. Both effects (porosity, micro-cracks) reduce the heat conductivity of the chondritic material, but in a different way.

\subsection{Dependence on porosity}

First we consider the effect of porosity. This is discussed in detail in \citet{Hen16} for material at room temperature. It was demonstrated that the effective porosity can be accounted for with reasonable accuracy by introducing in the Bruggeman mixing rule, Eq. (\ref{EffMediumBrugg}), the voids as an additional mixture component with volume fraction $f_i$ and vanishing heat conductivity $K_i$. The heat conductivities of the other components in the mixture are not modified by the presence of voids. Figure \ref{FigKTcomp}a shows as dashed lines the heat conductivity calculated in this way for the case of a pore space with a volume fraction of 10\% of the total volume and with spherical pores. This is within the range of porosities found for ordinary chondrites \citep{Con08}.

As to be expected, by the presence of pores the run of $K_\mathrm{eff}$ with temperature $T$ is simply shifted to lower values without changing the shape of the function $K_\mathrm{eff}(T)$. By comparing Figs.~\ref{FigKTcomp}a and \ref{FigKTvar}a,b or with Fig.~3 in \citet{Ope12} one easily sees that the shape of the function $K_\mathrm{eff}(T)$ as derived for a polycrystalline material is substantially different from what one observes for meteorites. The only exception is the calculated function $K_\mathrm{eff}(T)$ for EH chondrite material, which shows some similarity with the empirical data for the EH and EL chondrite shown in Fig.~\ref{FigKTvar}b. For all H, L  and LL meteorites the temperature peak of the $K_\mathrm{eff}(T)$ variation is very much broader and shifted to much higher temperatures than for the compacted materials shown in Fig.~\ref{FigKTcomp}a. From Fig.~\ref{FigKTvar}c one immediately recognizes, however, that the type of temperature variation of $K_\mathrm{eff}(T)$ may be obtained, if the role of phonon scattering by walls is significantly increased over its value for crystalline material for which the laboratory data of the mixture components used in our calculation are obtained.

\subsection{%
Dependence on boundary scattering}

The data used in the above considerations on the heat conductivity of minerals refer to specimens with sizes of the order of mm to cm as they are used in laboratory measurements and usually also to crystalline material without abundant impurities and internal phase boundaries, lattice defects or cracks. The value of the heat conductivity is determined by the scattering of phonons which is dominated in this case by the so called umklapp-process and by impurity scattering. The term $v/L$ in Eq.~(\ref{TauScatt}) describing the scattering of phonons at internal or external surfaces is of minor or no importance in that case. 

The material of chondrites, however, is composed of small crystallites where interfaces between different materials scatter phonons, and it contains a lot of impurities. Additionally the material of meteorites is pervaded by numerous micro-cracks \citep{Bry17} from impact shocks suffered by the material during its long history in near-surface layers of the parent body or by the final excavating shock, and also from thermal fatique by diurnal temperature cycling \cite{Del14}. Such micro-cracks may have mutual distances from the order of millimetres down to values as low as a few microns \citep{Str10}. They act as internal surfaces for phonon scattering which limit the mean free path of phonons and in this way lower the heat conductivity compared to the materials used in laboratory measurements. 

\begin{table*}
\caption{Results of the fitting procedure.}

\begin{tabular}{llccrrrrrrrc}  
\hline
\hline
\noalign{\smallskip}
Meteorite & Type &  Shock & Density & $\Phi_\mathrm{ex}$ & $\Phi_\mathrm{op}$ & $s$ & $\alpha$ & $\chi^2$ & $N$ & $\bar{\chi}^2$ & Note \\
\noalign{\smallskip}
\hline
\noalign{\smallskip}
Monroe          & H4  &    & 3.80 &  5.9 & 17.4 & 29.0 & 7.58 & 0.945  &  9 & 0.158  & a \\
Wellman         & H5  &    & 3.82 &  6.1 &  8.3 & 9.56 & 2.20 & 1.16   &  9 & 0.194  & a \\
Gilgoin Station & H5  & S4 & 3.81 &  5.0 &  5.9 & 15.0 & 13.3 & 0.900  &  9 & 0.150  & a \\
Gladstone       & H5  & S3 & 3.75 &  5.0 &  5.0 & 123. & 36.1 & 1.33   &  9 & 0.220  & a \\
Arapahoe        & L5  & S4 & 3.61 &  2.5 &  2.5 & 32.9 & 25.5 & 31.2   &  9 & 5.207  & a \\
Farmington      & L5  & S4 & 3.59 &  5.5 &  5.5 & 29.4 & 14.4 & 49.3   &  9 & 8.211  & a \\
Kunashak        & L6  & S4 & 3.59 &  5.2 &  5.2 & 137. & 38.9 & 4.47   &  9 & 0.744  & a \\
Bruderheim      & L6  & S4 & 3.60 &  8.0 & 16.8 & 28.3 & 38.9 & 2.97   &  9 & 0.494  & a \\
New Concord     & L6  & S4 & 3.60 &  9.2 & 13.1 & 57.6 & 63.1 & 2.68   &  9 & 0.445  & a \\
Leedey A        & L6  & S3 & 3.62 & 10.4 & 10.4 & 1912 & 50.3 & 2.37   &  9 & 0.395  & a \\
Leadey B        & L6  & S3 & 3.63 & 10.0 & 10.0 & 1412 & 47.8 & 2.65   &  9 & 0.442  & a \\
Y-74156         & H4  &    & 3.80 &  9.2 & 15.5 & 110. & 23.8 & 1.89   &  9 & 0.315  & a \\
Y-74647         & H4.5&    & 3.83 &  9.1 & 22.8 & 70.0 & 27.9 & 1.88   &  9 & 0.313  & a \\
ALH-77294       & H5  &    & 3.84 & 12.9 & 19.5 & 4.31 & 68.6 & 5.22   &  9 & 0.870  & a \\
ALH-77288       & H6  &    & 3.77 &  2.0 & 11.1 & 1.58 & 15.9 & 5.99   &  9 & 0.999  & a \\
Y-74191         & L3  &    & 3.60 & 10.3 & 10.3 & 144. & 15.8 & 2.98   &  9 & 0.498  & a \\
Y-75097         & L4  &    & 3.65 & 10.3 & 10.8 & 159. & 35.9 & 11.5   &  9 & 1.929  & a \\
MET-78003       & L6  &    & 3.61 &  7.8 & 7.8  & 92.9 & 17.6 & 6.12   &  9 & 1.019  & a \\
ALH-78251       & L6  &    & 3.70 & 13.2 & 15.7 & 3.09 & 83.5 & 6.82   &  9 & 1.136  & a \\
ALH-78103       & L6  &    & 3.70 & 13.4 & 19.3 & 20.8 & 63.7 & 3.13   &  9 & 0.520  & a \\
ALH-77231       & L6  &    & 3.59 & 14.3 & 21.6 & 156. & 19.4 & 3.30   &  9 & 0.550  & a \\
ALH-769         & L6  &    & 3.58 & 19.1 & 19.4 & 631. & 18.2 & 6.25   &  9 & 1.042  & a \\[.2cm]
Abee            & EH6 & S2-S5 & 3.62 &  3.0 &  3.5 & 1.22 & 58.8 & 1.194  &  51 & 0.0249 & b \\
Barbotan        & H5  & S3 & 3.75 &  6.9 &  6.9 & 22.9 & 17.2 & 3.254  &  54 & 0.0638 & b \\
Collescipoli    & H5  &    &      &      &  9.0 & 2290 & 7.94 & 0.801  &  39 & 0.0225 & b \\ 
Cronstadt       & H5  &    &      &      &  7.9 & 181. & 7.94 & 0.903  & 144 & 0.0064 & b \\
Bath Furnace 3  & L6  &    & 3.66 &  4.3 &  4.3 & 12.5 & 10.4 & 100.7  & 108 & 0.9595 & b,c \\
Lumpkin         & L6  &    &      &      &  7.3 & 14.5 & 65.2 & 0.716  &  45 & 0.0176 & b \\
Holbrook 2      & L6  & S2 & 3.55 & 10.4 & 19.3 & 279. & 44.1 & 0.275  &  24 & 0.0013 & b,c \\
La Cienega      & H6  &    &      &      & 10.0 & 6.31 & 63.8 & 4.829  &  65 & 0.0789 & b \\
Phillistfer     & EL6 &    & 3.70 &  2.4 & 3.5  & 1.00 & 53.7 & 17.39  & 105 & 0.1710 & b \\
Pultusk         & H5  & S3 & 3.72 &  7.5 & 9.1  & 573. & 31.6 & 0.379  &  57 & 0.0702 & b \\
\noalign{\smallskip}
\hline
\end{tabular}
\tablefoot{ References: (a) \citet{Yom83}, (b) \citet{Ope10,Ope12}. 

(c) From the measurements in different directions only one is used.
}

\label{TapFitMet}
\end{table*}

In our fit formula, Eq.~(\ref{ModelDebyeFit}), for the heat conductivity of crystalline material the average distance $L$ between subsequent scattering events of a phonon at the border or at internal surfaces is included in the parameter $p_1$. By reducing the average distance from some value $L$ to a smaller value $l$ increases the value of the parameter $p_1$ to the new value $p_1(L/l)$. Hence, for calculating the heat conductivity of a specimen with enhanced scattering at internal surfaces we have to scale the parameter $p_1$ by an appropriate factor $s=L/l>1$. The possibly high density of micro cracks may require a scaling of $L$ from the cm scale to the micrometer scale (scaling factors up to $s\approx10^3$),

In Fig.~\ref{FigKTcomp}b the dashed and the dashed-dotted lines shows the heat conductivity calculated for chondritic material where the typical distance $L$ between reflecting walls is reduced by factors of $s=10$, $s=100$, and $s=1000$, respectively. Two effects are obvious: First, a significant reduction of the heat conductivity and, second, a reduction of the temperature variation of the heat conductivity at high temperatures because of the temperature independence of wall-scattering. The maximum of the temperature variation of $K_\mathrm{eff}(T)$ is shifted to higher temperatures or the variation of  $K_\mathrm{eff}(T)$ is flattened to such an extent that a maximum is hardly to recognize in the plotted temperature range. In order to facilitate the comparison of the temperature variation of the different curves, Fig.~\ref{FigKTcomp}c shows them normalized to their maximum value. The kind of temperature variation of $K_\mathrm{eff}(T)/K_\mathrm{max}$ obtained here strongly resembles what one sees in  Fig.~\ref{FigKTvar}a,b for the H, L, and LL chondrites. The different way by which the variation of the heat conductivity is modified by macroscopic porosity and micro-cracks results from the following: Pores increase the path-length for the heat flow in a material, which reduces heat conductivity in a temperature-independent way. Micro-cracks change the relative contribution of the temperature-independent surface or interface scattering processes relative to the temperature-dependent processes of impurity- and umklapp-process scattering to the relaxation time $\tau$ for phonon scattering. This results in temperature-dependent reductions of the heat conductivity. 

Our results suggest that the phonon scattering in meteorites is strongly enhanced compared, e.g., to the polymineralic and polycrystalline terrestrial rock materials which show temperature variations of  $K_\mathrm{eff}(T)$ resembling the full lines in Fig.~\ref{FigKTcomp}b. The observed heat conductivity of the material of a meteorite, thus, is to large extent determined by the impact history of the surface region of the parent body from which a specific meteorite is excavated, which determines the density of micro-cracks in the material.

\subsection{%
Comparison with some special meteorites}

Two sets of measured heat conductivities of chondrites at different temperatures are available in the literature. The study of \citet{Yom83} covers the temperature range between 100 K and 500 K. The studies of \citet{Ope10} and \citet{Ope12} consider the conductivity at low-temperatures between $\sim5$ K and 300 K. The first set of data is applicable to the problem of thermal evolution of planetesimals during the early evolution of the solar system. Though higher temperatures up to the melting temperatures of the main components would also be required for such applications, at temperatures above about 600 K beyond which radiative transfer starts to contribute to the total heat conductivity which means that such data would not be suited for our present purpose. The data set of \citet{Ope10,Ope12} is more applicable to calculate temperatures of bodies in the present-day asteroid belt. In the following we attempt to fit our model for the temperature dependent heat conductivity to laboratory-determined data for the variation of heat conductivity of the meteorites of these two samples. In this way we check if the model is able to reproduce the thermal properties of meteoritic material. 

\begin{figure*}

\includegraphics[width=.33\hsize]{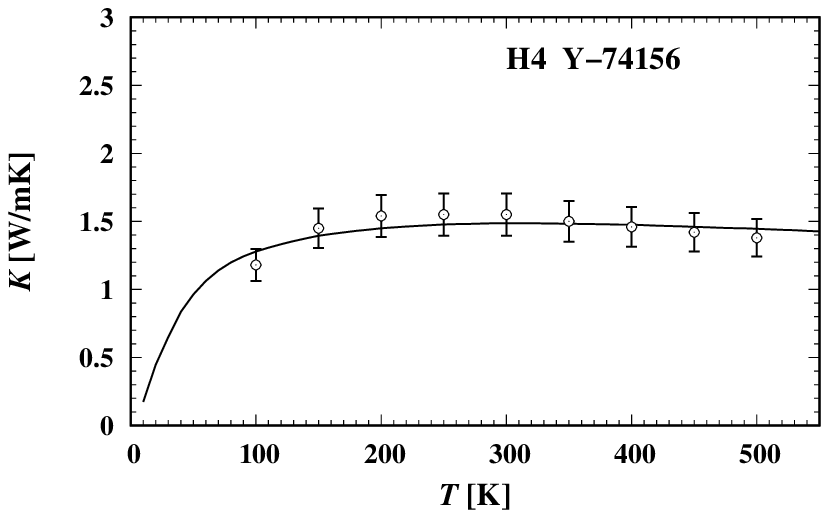}
\includegraphics[width=.33\hsize]{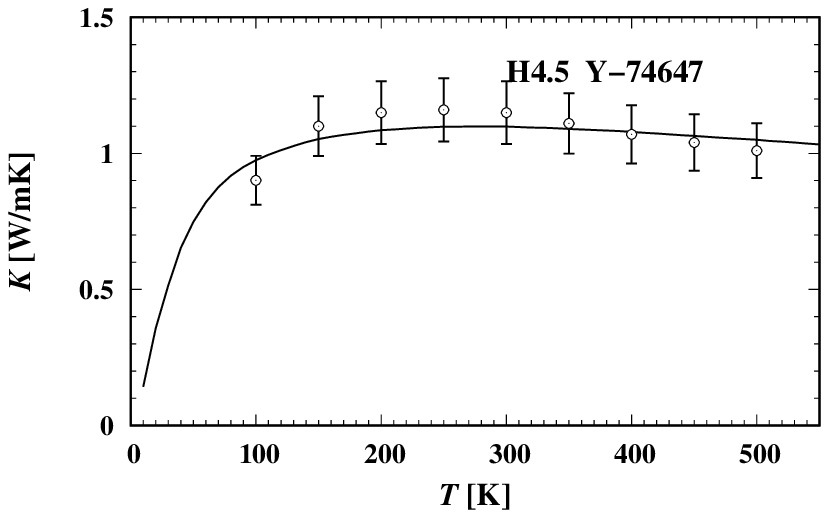}
\includegraphics[width=.33\hsize]{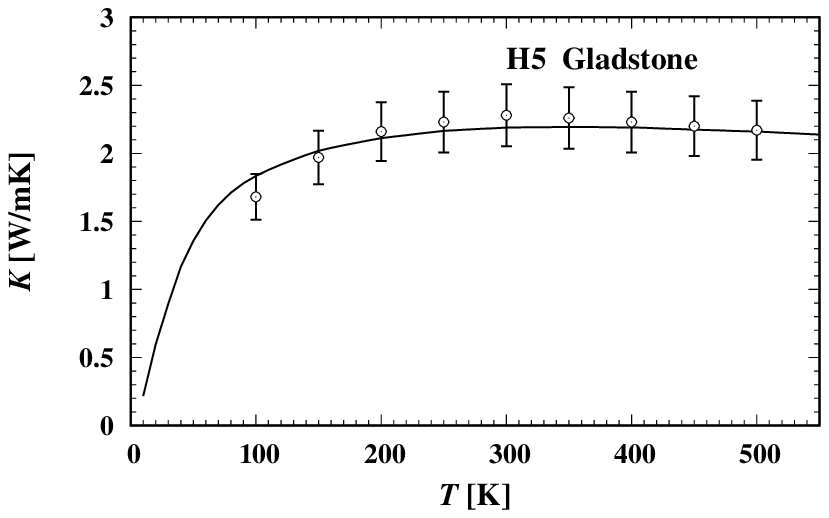}

\includegraphics[width=.33\hsize]{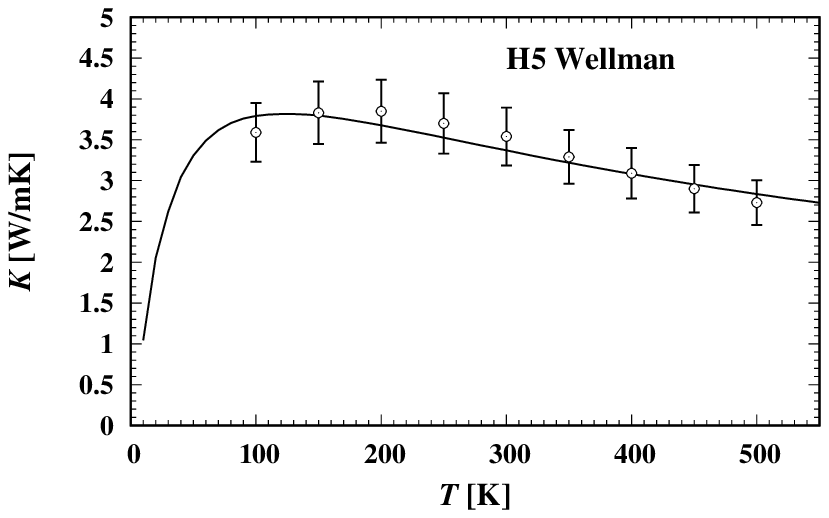}
\includegraphics[width=.33\hsize]{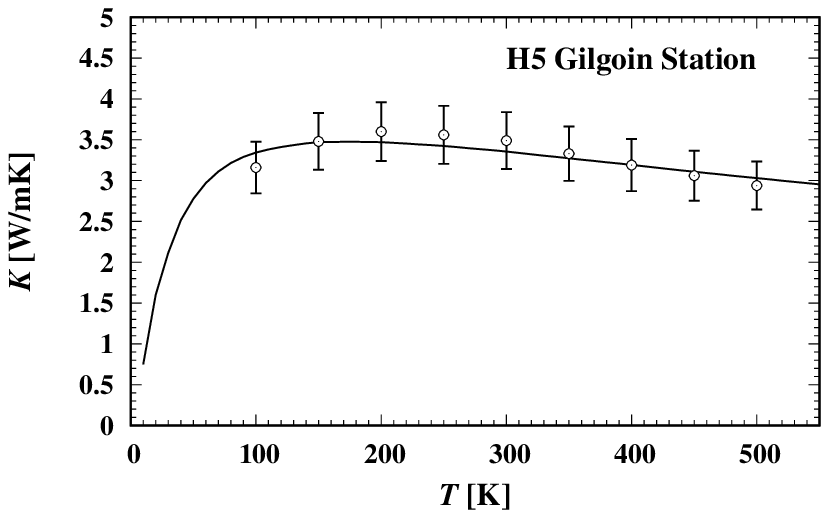}
\includegraphics[width=.33\hsize]{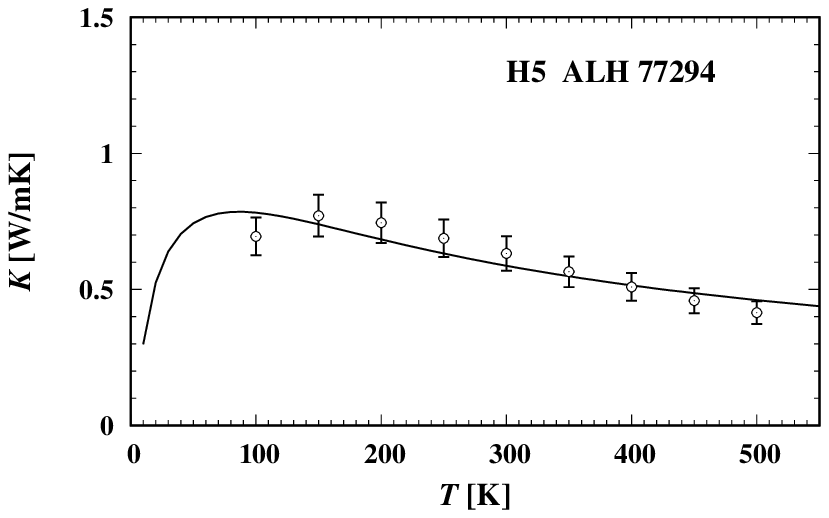}

\includegraphics[width=.33\hsize]{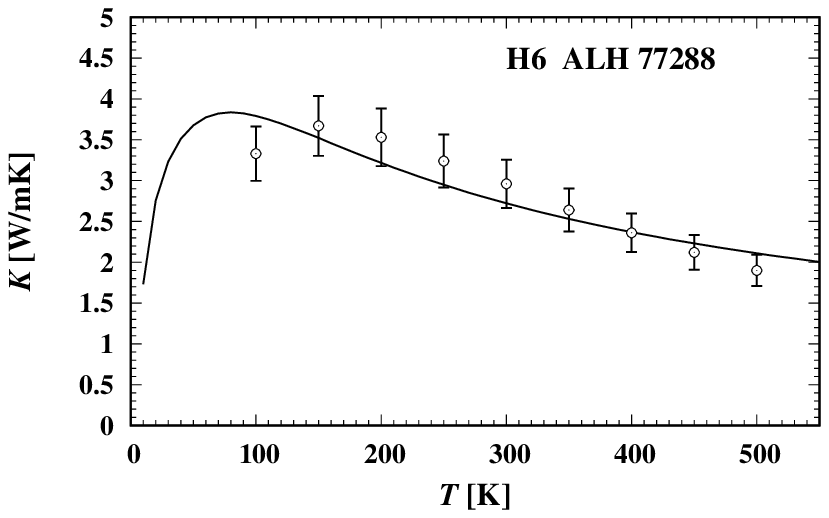}
\includegraphics[width=.33\hsize]{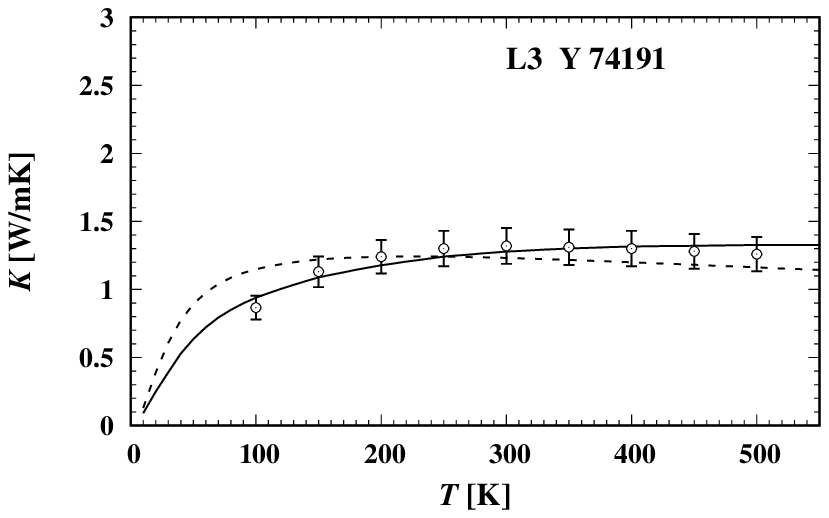}
\includegraphics[width=.33\hsize]{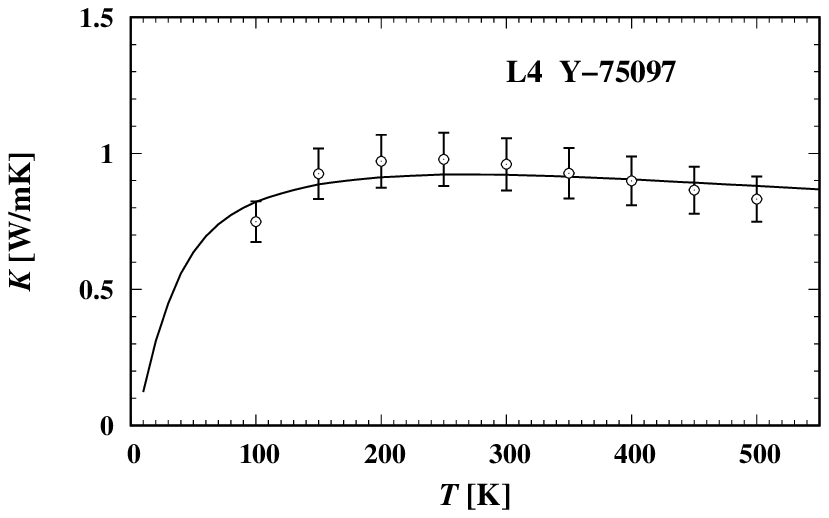}

\includegraphics[width=.33\hsize]{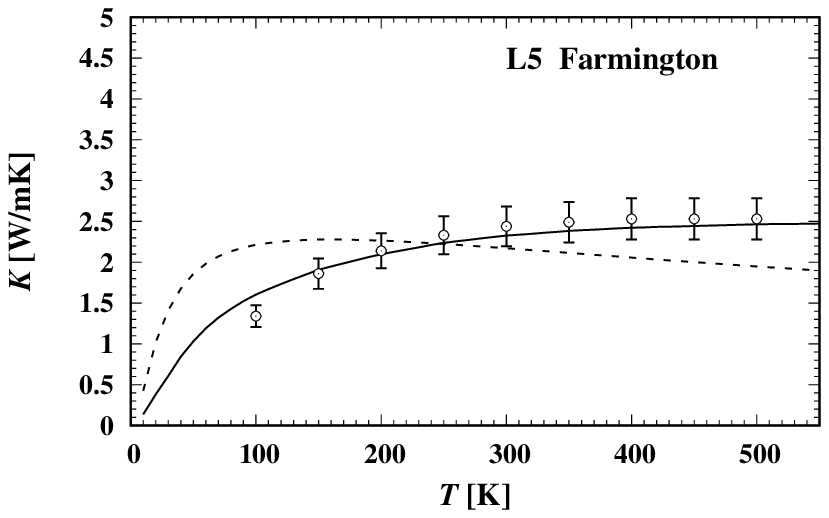}
\includegraphics[width=.33\hsize]{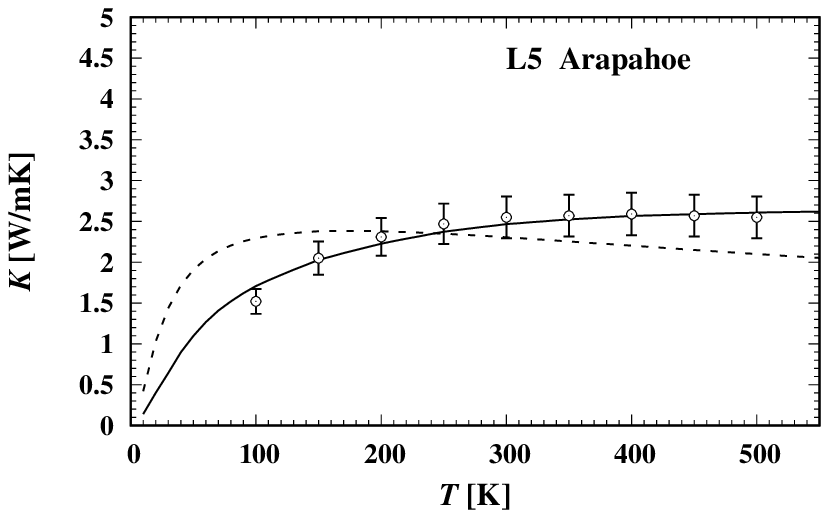}
\includegraphics[width=.33\hsize]{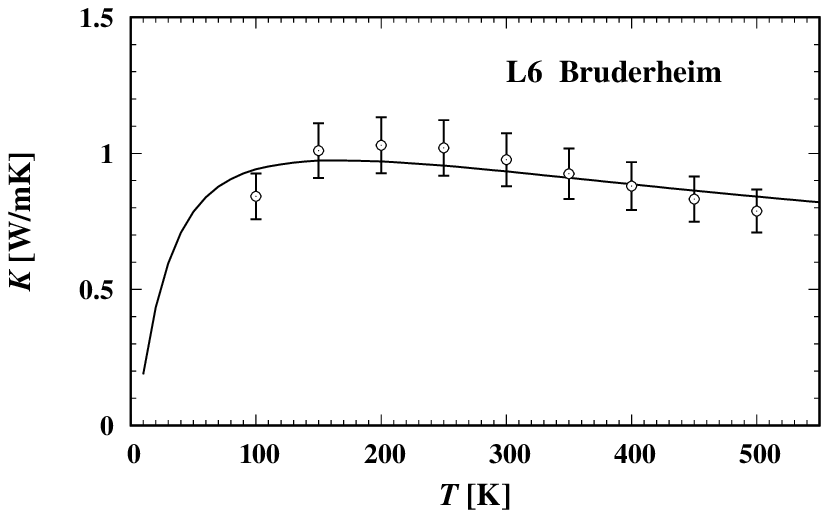}

\includegraphics[width=.33\hsize]{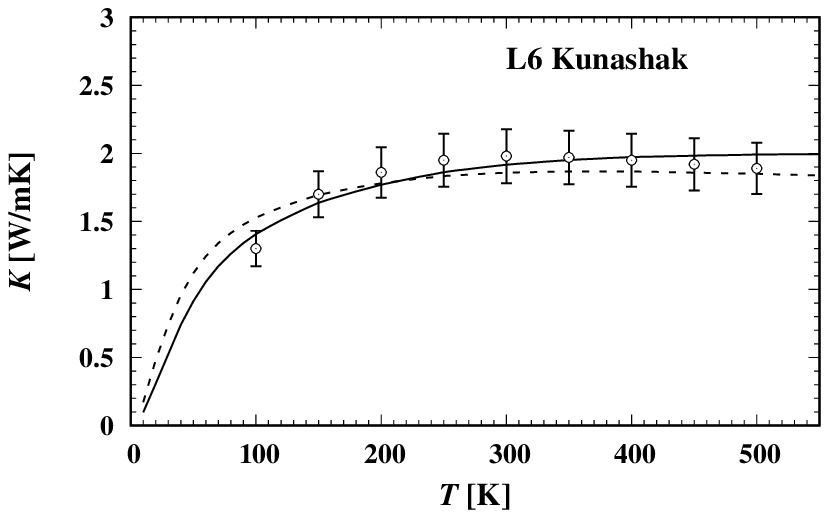}
\includegraphics[width=.33\hsize]{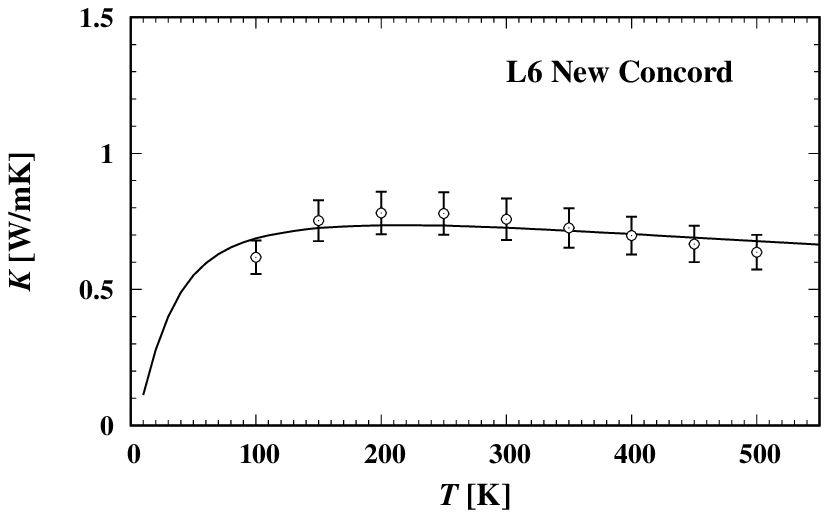}
\includegraphics[width=.33\hsize]{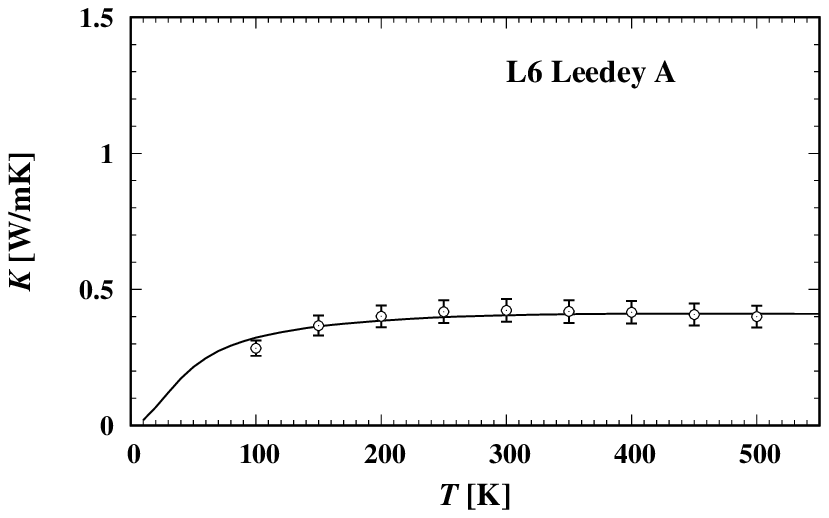}

\caption{Comparison of experimental data for heat conductivity of meteorites with the result of modelling. Circles with error bars represent the data from \citet{Yom83} with their estimated error. The solid line represents the optimum model fit. The dashed line shows the model fit with fixed composition in the case were this results in no acceptable fit. In the latter case the solid line represents a different fit with variable iron volume fraction.}

\label{FigOptYom}
\end{figure*}

We calculate the effective heat conductivity $K_\mathrm{eff}$ of the meteorites for simplicity by means of the Bruggeman mixing rule. In \citet{Hen16} we found, that this yields almost the same results compared to a direct numerical solution of the heat conduction equation using arbitrarily shaped sub-units of a multi-component mixed material. The mixing rule is \citep[cf.][]{Ber95} 
\begin{equation}
\sum\limits_{i=1}^Nf_n\left(K_i-K_\mathrm{eff}\right){1\over9}\sum_{j=1}^3{1\over L_jK_i+(1-L_j)K_\mathrm{eff}}=0\,,
\label{EffMediumBrugEllip}
\end{equation}
which is the generalised version of Eq.~(\ref{EffMediumBrugg}) if the sub-units are not spherical. The sum over $i$ in this expression runs over all components in the mixture (solid components and voids in our case), $f_i$ are their volume fractions, and $K_i$ their heat conductivities. The $L_j$ are the depolarisation coefficients of the non-spherical inclusions which are assumed to be ellipsoids. They derive their name from the analogous electrostatic problem where they describe the modification of the electric field strength inside an inclusion by influenced charges at the boundary layer between materials with deviating dielectric constant $\varepsilon$. The sum $j$ runs over the principal axis of the ellipsoids.

\begin{figure*}

\addtocounter{figure}{-1}

\includegraphics[width=.33\hsize]{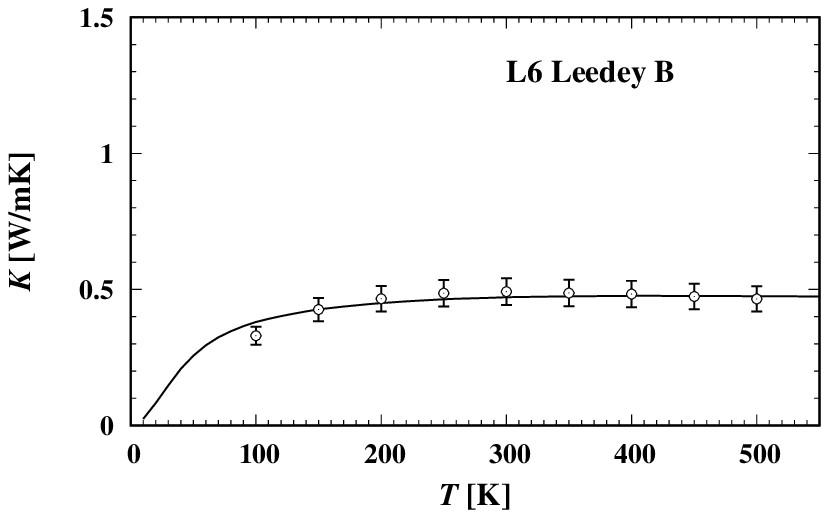}
\includegraphics[width=.33\hsize]{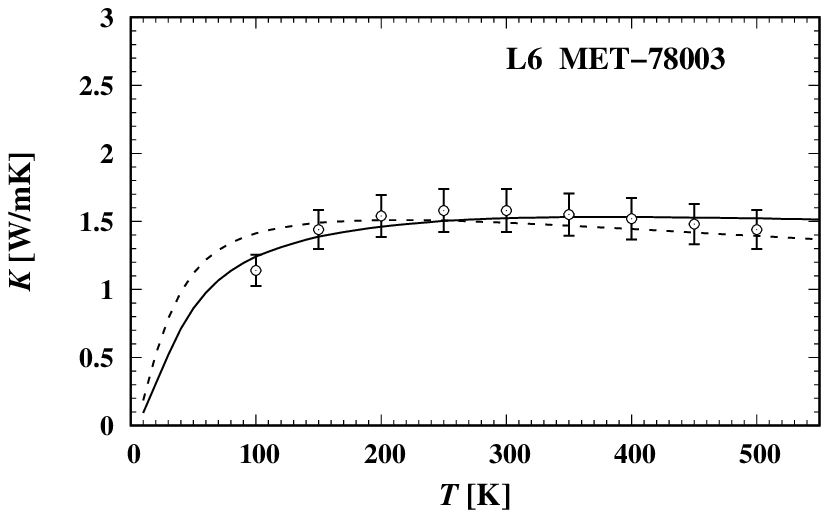}
\includegraphics[width=.33\hsize]{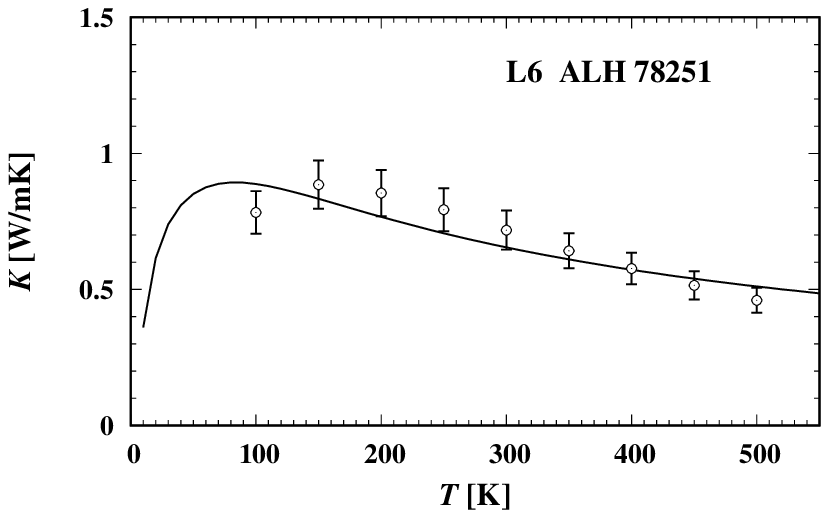}

\includegraphics[width=.33\hsize]{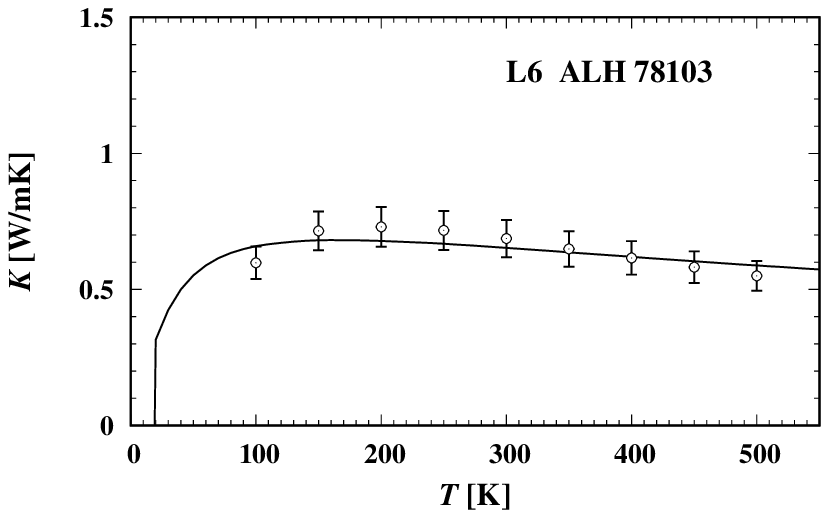}
\includegraphics[width=.33\hsize]{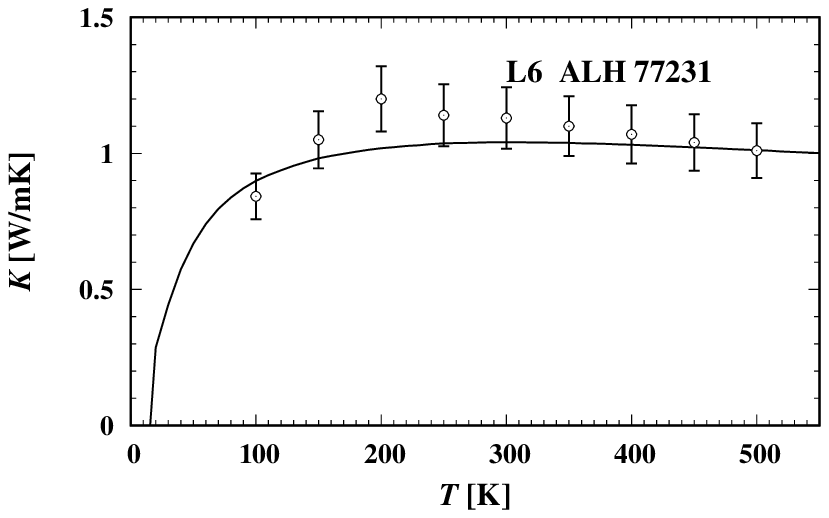}
\includegraphics[width=.33\hsize]{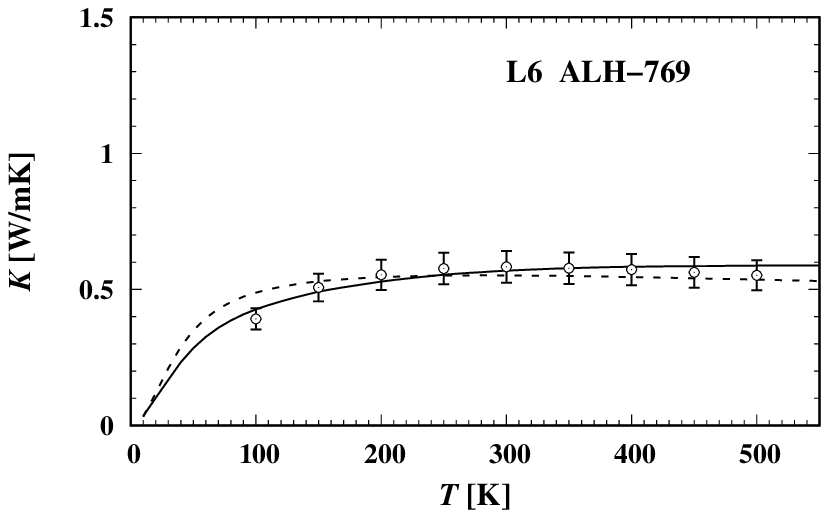}

\caption{Continued. }

\end{figure*}

The depolarisation coefficients are for an oblate rotational ellipsoid with principal axis $A=B>C$ are \citep{Boh83}
\begin{equation}
L_1={g\over2e^2}\left({\pi\over2}-\arctan g\right)-{g^2\over2}\,, 
\label{Depolar}
\end{equation}
with $L_2=L_1$ and $L_1+L_2+L_3=1$, where
\begin{equation}
e^2=1-{C^2/A^2}\,,\quad g={\sqrt{1-e^2}\over e}\,.
\end{equation}

Equation (\ref{EffMediumBrugEllip}) is a non-linear equation for $K_\mathrm{eff}$ which has to be solved numerically, which is done by Newton-Raphson iteration. The $K_i$ are calculated according to Eq.~(\ref{ModelDebyeFit}) by a numerical integration using the coefficients given in Table~\ref{TabVarParm1}. The value of $K_i$ for nickel-iron is determined by interpolation in the table of \citet{Ho78}.

The comparison which we plan to do cannot be done straight forward, however, since some properties of the meteoritic material are not known. The most important one is the possible reduction of the scattering length of phonons by abundant micro-cracks resulting from impact shocks. We simulate this by multiplying the parameter $p_1$  in Eq.~(\ref{ModelDebyeFit}) by a scaling factor $s>1$. The other one is the aspect ratio $\alpha=A/C$ of the micro-cracks and that of the voids which are usually found in meteorites \citep[see][ for a review]{Con08}. The presence of voids and their aspect ratio has a significant influence on the heat conductivity \citep{Yom83,Hen16}. We consider both quantities, $s$ and $\alpha$, as free parameters and perform a least square fit of the model to the measurements, which fixes the values of these unknown quantities by optimisation. The optimisation is performed by means of a genetic algorithm in the special version described by \citet{Cha95}.

\subsubsection{The sample of Yomogida and Matsui}

We begin with the sample of H and L chondrites of \citet{Yom83}, which is listed in Table~\ref{TapFitMet}. This comprises 22 meteorites of type H and L, a few of petrologic type~3 and 4, but most of them of type~5 and 6. Meteorites of this high petrologic type are already significantly compacted such that the pore volume usually is small and their mineral inventory is crystalline. For the heat conductivity of the mineral components we then can use the conductivity model derived in Sect.~\ref{SectKComp}. 

For the purpose of calculating $K_\mathrm{eff}$ we additionally have to specify the composition and the porosity of the meteorites.

With respect to the composition we use for all meteorites the average composition of the corresponding meteorite class from Table~\ref{TabMetComp}. In a few cases, however, the iron content will be treated as an adjustable parameter for reasons that are explained later. The void space fraction $f_i$ corresponding to the porosity, $\Phi$, of the meteorites is treated as an additional mixture component with vanishing heat conductivity. A value for the porosity, $\Phi_\mathrm{ex}$ has been determined by \citet{Yom83} for each of the meteorites of their sample. We prefer, however, not to use this value for our model calculation since (i) the accuracy of their porosity determination is given as only several percent which may introduce noticeable errors in the value of the computed heat conductivity and (ii) the micro-cracks may have to some fraction no direct connection to the surface such that they are not registered as void space by the method of helium-pyknometry used by \citet{Yom83}. Instead, we consider $\Phi$ also as a free parameter that is determined by the optimisation procedure to  fit the model to the measured data, but take care of that $\Phi$ is not less than the measured value if one has been determined.

\begin{figure*}

\includegraphics[width=.33\hsize]{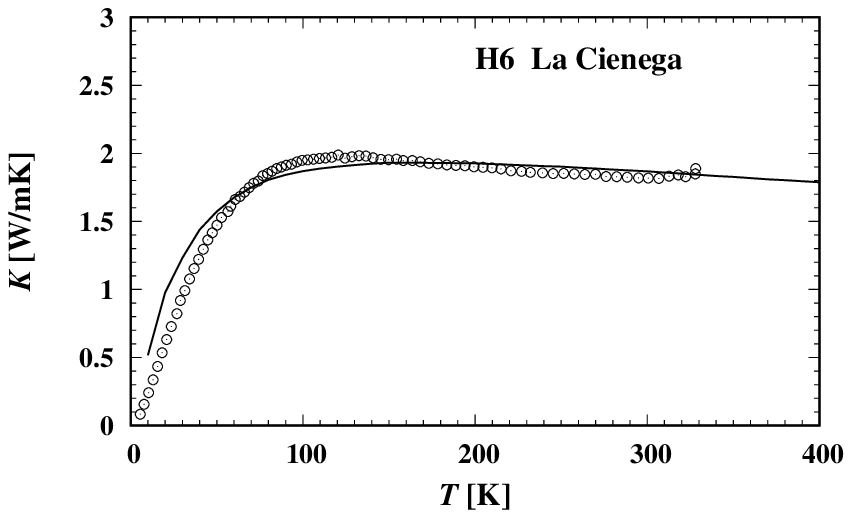}
\includegraphics[width=.33\hsize]{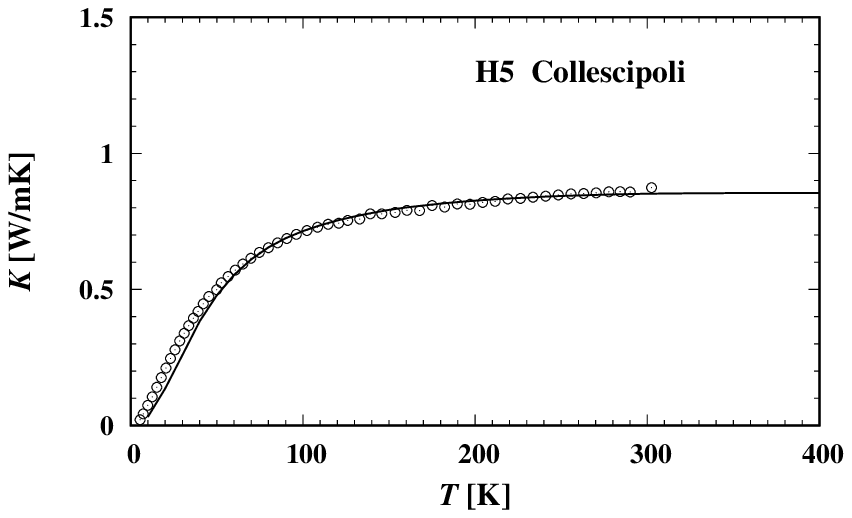}
\includegraphics[width=.33\hsize]{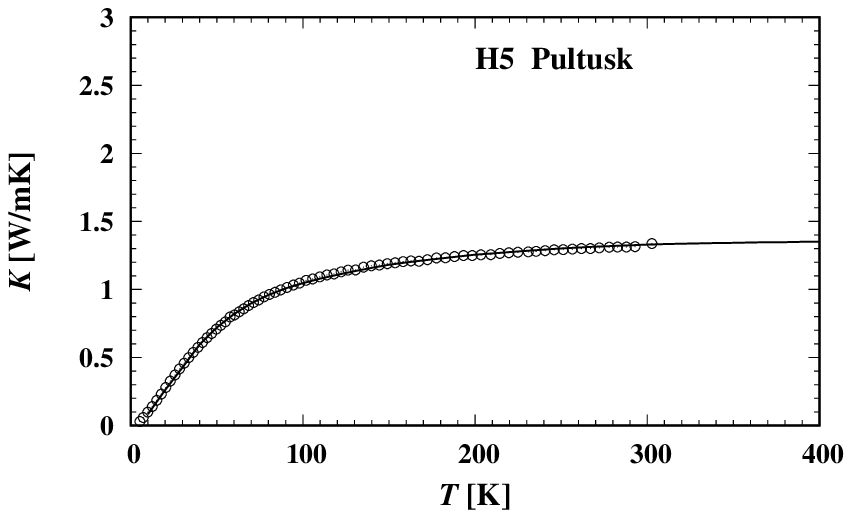}

\includegraphics[width=.33\hsize]{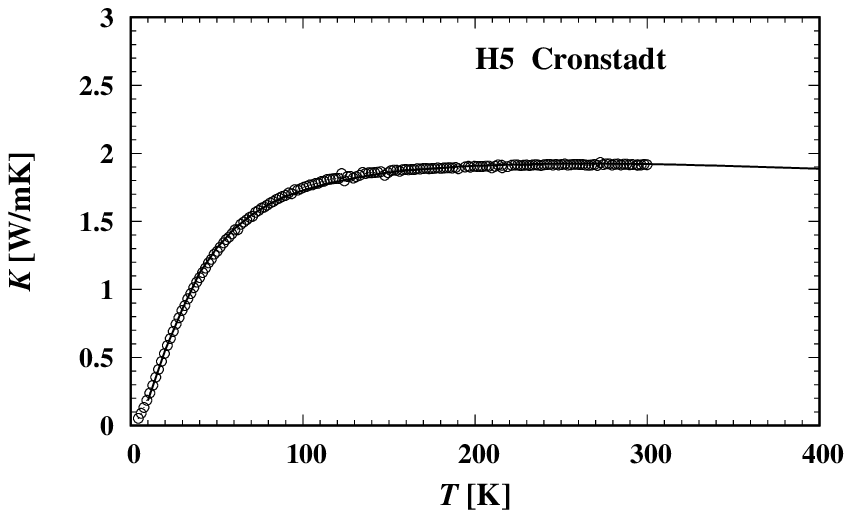}
\includegraphics[width=.33\hsize]{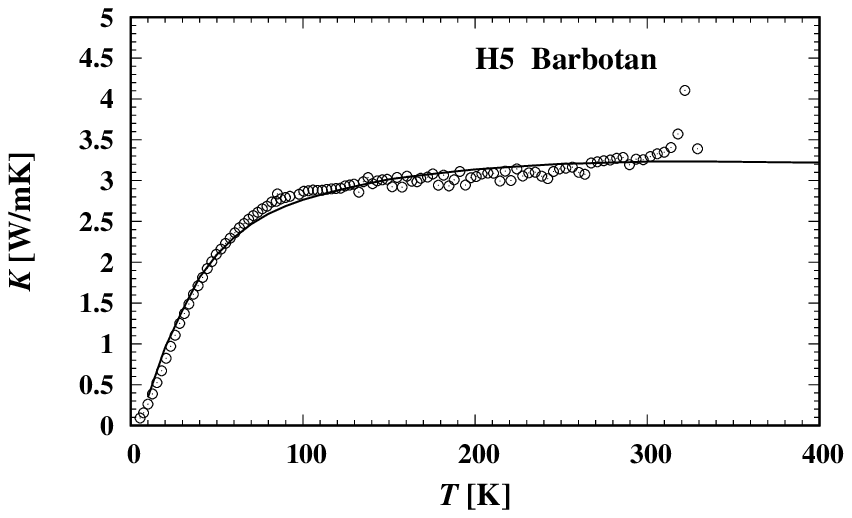}
\includegraphics[width=.33\hsize]{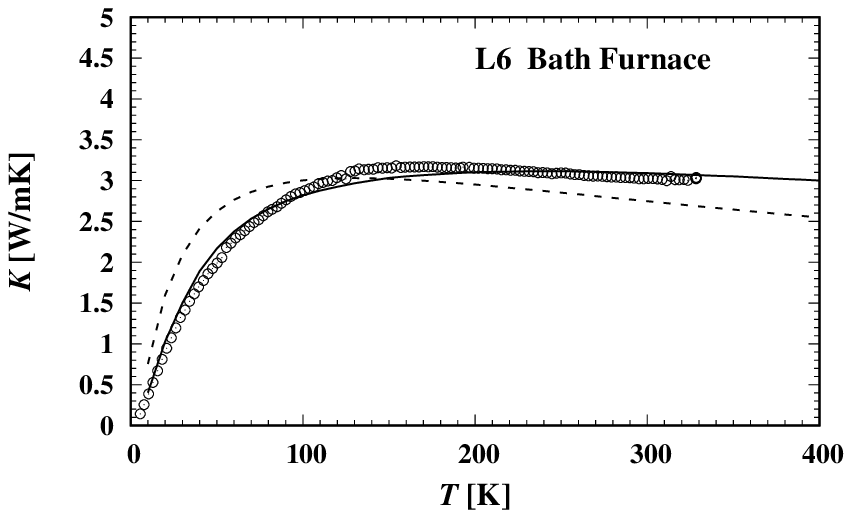}

\includegraphics[width=.33\hsize]{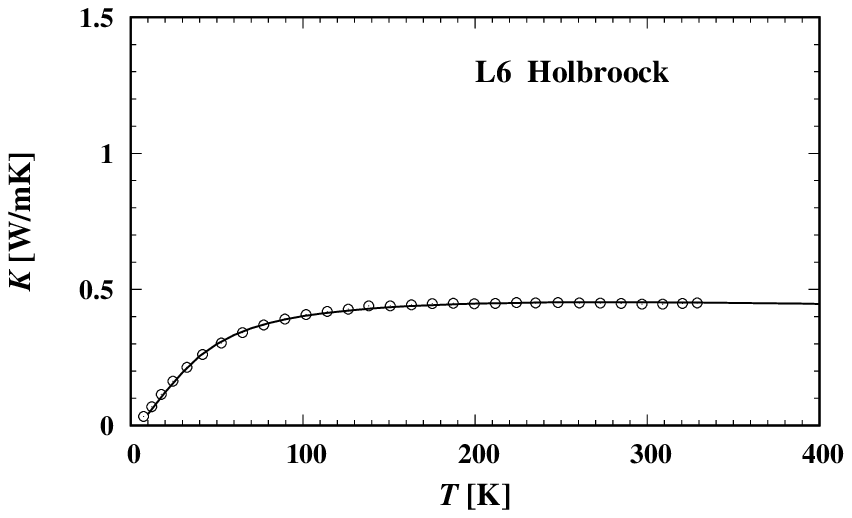}
\includegraphics[width=.33\hsize]{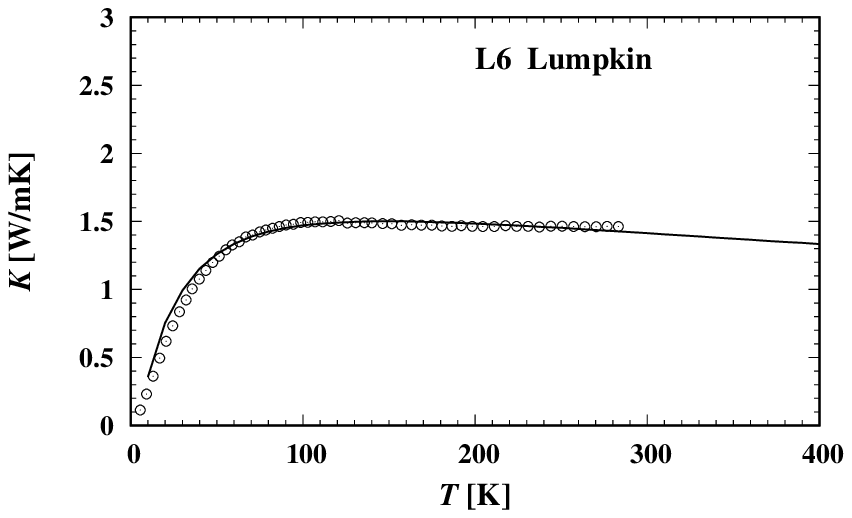}
\includegraphics[width=.33\hsize]{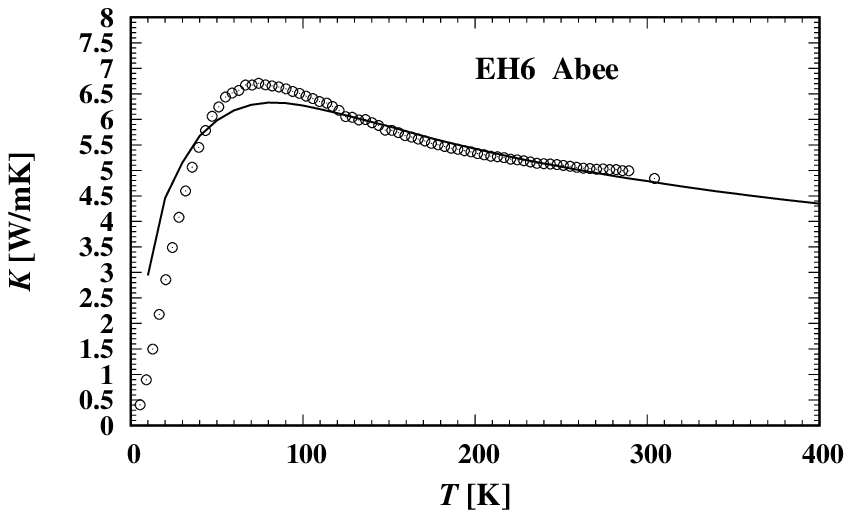}

\includegraphics[width=.33\hsize]{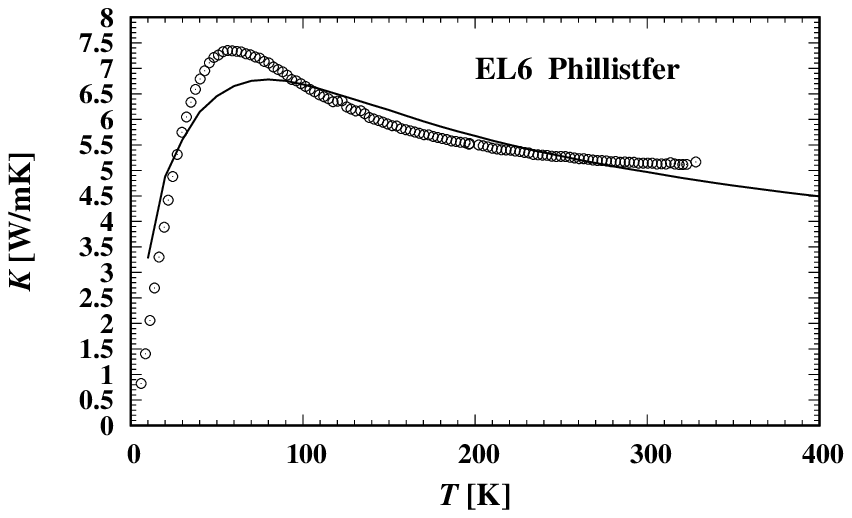}

\caption{Comparison of experimental data for heat conductivity of meteorites and the result of modelling. Circles represent the data from \citet{Ope10,Ope12}. The solid line represents the optimum model fit. The dashed line shows the model fit with fixed composition in the case were this results in no acceptable fit. In the latter case the solid line represents a different fit with variable iron volume fraction.}

\label{FigOptOpeil}
\end{figure*}

For the experimental heat conductivity data we use the values given in Table~8 of \citet{Yom83}. These are obviously not the original experimental data but already values determined from their fit to the experimental data. For our purposes this should suffice, since the accuracy is of the data of \citet{Yom83} is estimated by them to be only $\pm10$\%. We have determined fits of the model to the data for all of their H and L chondrites by minimising the quality function
\begin{equation}
\chi^2=\sum\limits_{n=1}^N\left[{K(T_n)-K_n\over\sigma_n}\right]^2\,.
\label{DefChi2}
\end{equation}  
This corresponds to the least mean square, defined as the global minimum value of $\chi^2$ for any parameter combination $\Phi$, $s$, $\alpha$. Here, $K_n$ is the experimental value at temperature $T_n$, $K(T_n)$ the corresponding value calculated by the model using the same temperature and a set of parameters $\Phi$, $s$, $\alpha$, and the sum over $n$ runs over the $N$ available data for a meteorite. The quantity $\sigma_n$ is the estimated error of the experimental data. The optimisation is done with the genetic algorithm \citep{Cha95} for at least 50 generations using 100 individuals per generation. For each meteorite a couple of optimisations are run using different seeds for the random number generator.

The accuracy of the empirical data is given by \citet{Yom83} as about $\pm10$\%. We use in Eq.~(\ref{DefChi2}) the following value for $\sigma_n$:
\begin{equation}
\sigma_n=0.1\times K_n\,.
\label{DefSigForFit}
\end{equation}
A reliable fit requires 
\begin{equation}
\bar{\chi}^2=\chi^2/(N-P)<1\,.
\end{equation}
The number of data as given by \citet{Yom83} is $N=9$ for all their meteorites. The number of parameters $P$ with respect to which a minimisation is performed is $P=3$. 

Table~\ref{TapFitMet} shows the values of the parameters $s$, $\alpha$, and $\Phi_\mathrm{op}$ corresponding to the optimum fit between the data and the model, and Fig.~\ref{FigOptYom} shows the optimum fits for the individual meteorites compared to the laboratory data. Table~\ref{TapFitMet} also shows the value of $\bar{\chi}^2$ which decides on the reliability of the fit. An inspection of the table and the figure shows, that in most cases it is possible to obtain a reliable fit with $\bar\chi^2<1$. Our model therefore is able to explain the observed temperature variation of the heat conductivity, $K$, of chondritic material.

\begin{table}
\caption{Models for meteorites from the sample of \citet{Yom83} and \citet{Ope12} with variable iron volume fraction $f_\mathrm{ir}$.}

\begin{tabular}{@{}lrrrrll@{}}
\hline
\hline
\noalign{\smallskip}
Meteorite & $\Phi_\mathrm{op}$ & \multicolumn{1}{c}{$s$} & \multicolumn{1}{c}{$\alpha$} & \multicolumn{1}{c}{$f_\mathrm{ir}$} & \multicolumn{1}{c}{$\chi^2$} & \multicolumn{1}{c}{$\bar{\chi}^2$}  \\
\noalign{\smallskip}
\hline
\noalign{\smallskip}
Y-74191        & 17.3 & 307. & 36.7 & 0.157 & 1.753 & 0.292 \\ 
Farmington     &  5.5 & 496. & 88.0 & 0.245 & 4.250 & 0.708 \\
Kunashak       & 15.1 & 1000 & 11.3 & 0.227 & 1.726 & 0.288 \\
Arapahoe       &  3.3 & 543. & 122. & 0.245 & 1.854 & 0.371 \\
MET-78003      & 15.8 & 532. & 7.23 & 0.163 & 1.932 & 0.386 \\
ALH-769        & 26.9 & 865. & 34.7 & 0.118 & 1.779 & 0.356 \\[.2cm]
Bath Furnace 3 &  6.5 & 62.9 & 2.00 & 0.203 & 8.684 & 0.083 \\ 
\noalign{\smallskip}
\hline
\end{tabular}

\label{TapFitMetVarFe}
\end{table}

For a number cases, however, the best value of $\bar{\chi}^2$ is definitively much bigger than one and the fit cannot be considered as reliable. In these cases the corresponding fit is shown in Fig.~\ref{FigOptYom} with a dashed line. This problem may result from the fact that the samples used for thermal diffusivity measurements by \citet{Yom83} were rather small rectangular parallelepipeds of about 3.5 \dots\ 4 mm side length. Since the iron component of the material in H and L chondrites is found as irregular shaped and often strongly elongated particulates with sizes mostly $\ll 1$~mm, but in part with the longest edge having sizes of the order of up to $>1$ mm, and because they are spatially inhomogeneously distributed on the mm-scale, there is the possibility that for the rather small samples used the volume fraction of iron in an individual sample may be somewhat higher than the average. Because of the high conductivity of iron this may result in a noticeable modification of the average heat conductivity of the sample. For this reason we consider for some of the meteorites, for which no good fit can be obtained with the average composition, the volume fraction of iron as a free parameter which is also determined via the optimisation process.  

The results for such a fit are shown in Table~\ref{TapFitMetVarFe}. In all cases the value of $\bar{\chi}^2$ is significantly reduced and an inspection of Fig.~\ref{FigOptYom} shows that the quality of the new fit is much improved. The volume fraction of iron required for the improvement appears somewhat high, however, such that it is not clear if this is the right explanation. An alternative cause for the low quality of the fit in some cases may be that the data published in \citet{Yom83} seem to be calculated from their fit to the measured values and may already be somewhat misleading because a special functional form is assumed for the fit that may not be suited for every physically possible case.

We tested also for the other meteorites if a variable iron content would result in an improved fit quality, but found only small but no significant reductions of $\bar{\chi}^2$ in those cases. 

\subsubsection{The sample of Opeil and Consolmagno}

Here we consider the sample of H, L, and E chondrites of \citet{Ope10,Ope12} for which very detailed reults for the temperature variation of heat conductivity at low temperatures were obtained. The original data were kindly made available for the present study by G. Consolmagno S.J. Here we consider the meteorites listed in Table~\ref{TapFitMet}.

The general procedure is as in the preceding case. The temperature range of the data extends here to very low temperatures. It is unclear wether our heat conductivity model for the silicate components is of sufficient accuracy in the low-temperature range $\lesssim100$ K because, for lack of experimental data for the low-temperature conductivity of the mineral components, the Debye temperature $\Theta_\mathrm{D}$ could only be determined by the approximate relation Eq.~(\ref{RelDebTemps}). Therefore we exclude the temperature points $T<50$ K of the laboratory data from the calculation of $\chi^2$. 

The parameter values $\Phi$, $s$, and $\alpha$ corresponding to the minimum of $\chi^2$ are shown in Table~\ref{TapFitMet} and the model fit is compared with the data in Fig.~\ref{FigOptOpeil}. We obtain for the H and L meteorites very good fits, except for Bath Furnace, even at the lowest temperatures which are in principle outside the range of applicability of our model. If in the case of Bath furnace the volume fraction of metal is considered as as free parameter, the quality of the fit becomes quite good, see Tab.~\ref{TapFitMetVarFe}. 

Only for the two enstatite chondrites there are significant deviations in the low-temperature range. The method applied here to estimate the parameter values of the Callaway model of heat conductivity of crystalline material obviously does not provide a good fit for enstatite. For our calculation of the heat conductivity of H and L chondrites this has no strong implications because the pyroxene component represents only a small fraction of the total silicate material of the meteorite, while for enstatite chondrites it strongly dominates and is responsible for the discrepancy seen at temperatures below 100 K. 
 
The low values $\bar\chi^2\ll1$ for the meteorites in the present sample compared to the previous sample is suspicious. This results to large extent from using for reasons of comparability in the present case for $\sigma_n$ also a value as given by Eq.~(\ref{DefSigForFit}) which is adequate for the accuracy of the experimental data of \citep{Yom83}. For the accuracy of the their measurment \citet{Ope12} give, however, a value of $\sim1$\%. Using this for calculating $\bar\chi^2$ would bring the resulting new value of $\bar\chi^2$ close to $\bar\chi^2\lesssim1$ similar as in the preceding case.

\begin{figure}

\includegraphics[width=\hsize]{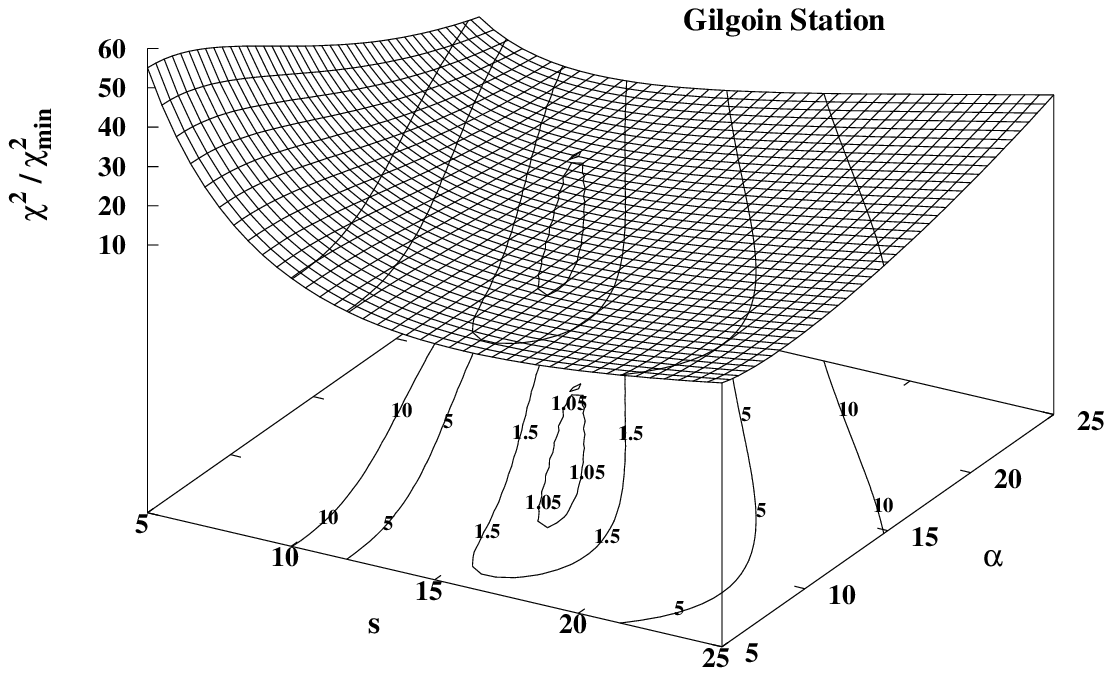}

\includegraphics[width=\hsize]{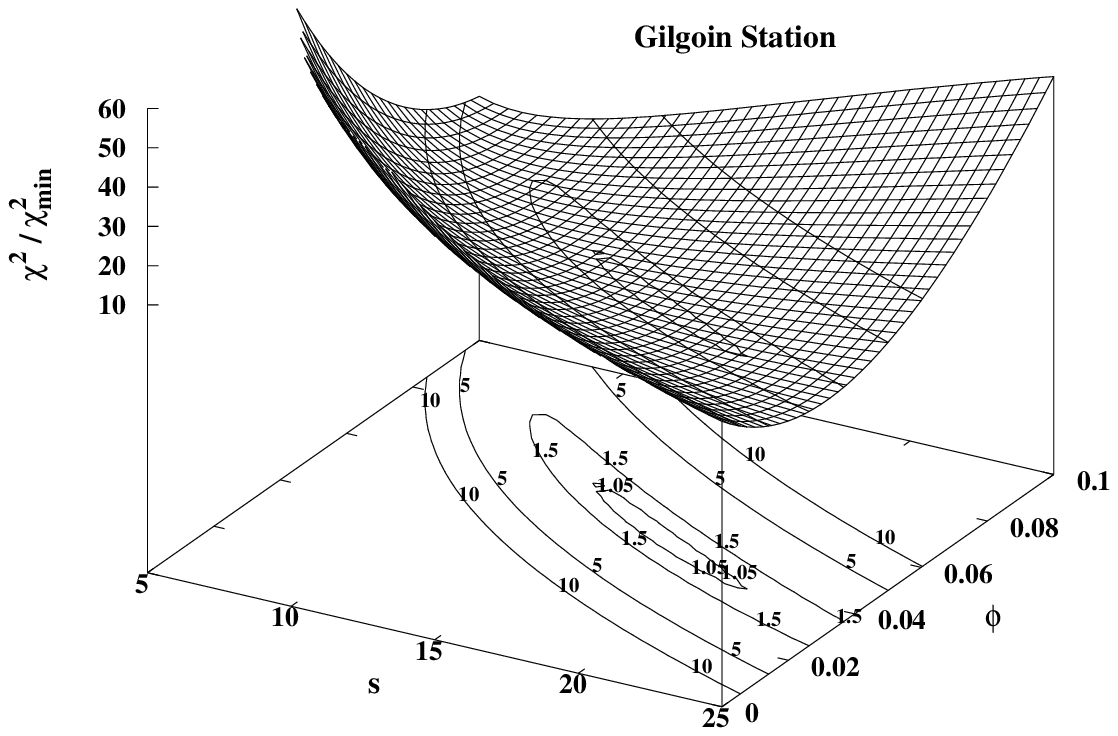}

\includegraphics[width=\hsize]{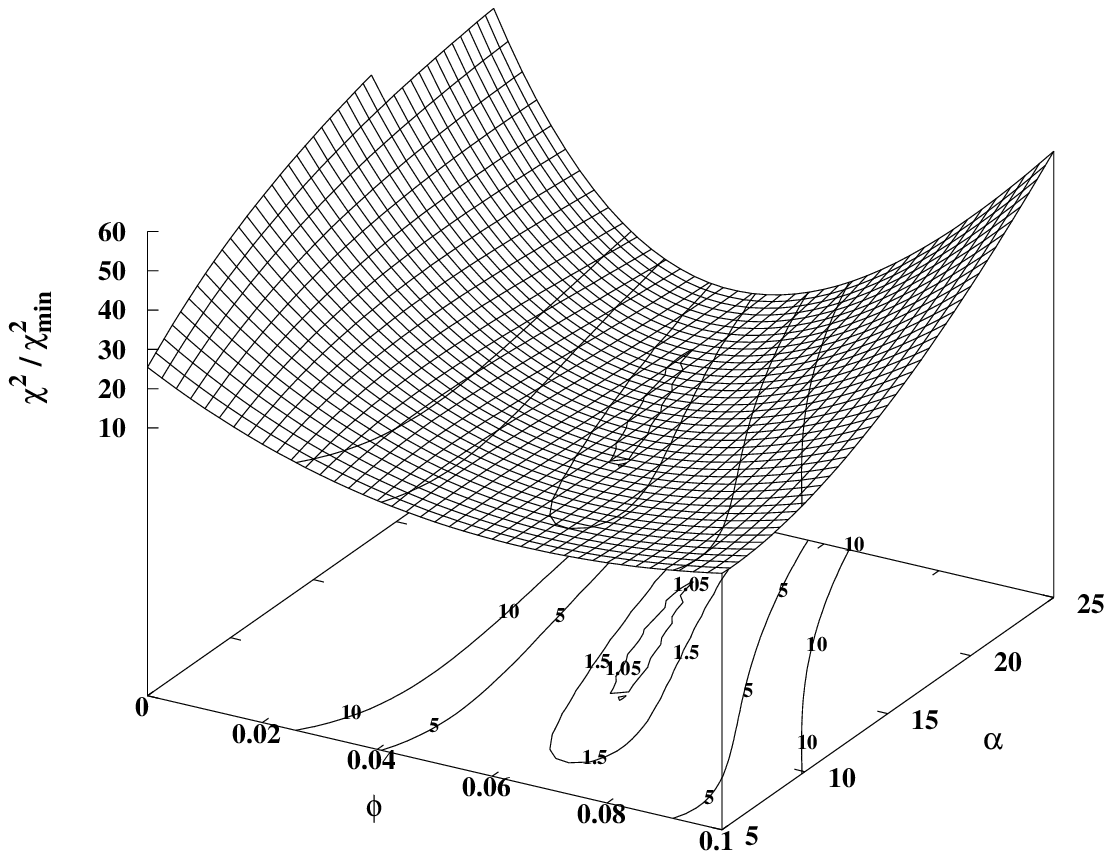}

\caption{Variation of normalised $\chi^2$ with parameters $s$ and $\alpha$ for fixed $\Phi$ (upper panel), with parameters $s$ and $\Phi$ for fixed $\alpha$ (middle panel), and with parameters $\Phi$ and $\alpha$ for fixed $s$ (lower panel). Also shown are the contour lines for $\chi^2/\chi^2_\mathrm{min}=1.05$, 1.5, 5, and 10 and their projection onto the basis plane.}

\label{FigDepPar}
\end{figure}

\subsubsection{Sensitivity of the fit}

The dependency of the fit quality on the parameter values $s$, $\alpha$ , and $\Phi$ is shown in Fig.~\ref{FigDepPar} where the ratio of $\chi^2$ to the minimum value $\chi^2_\mathrm{min}$ found by the optimisation is plotted in some range of the parameter values around the optimum-fit values of $s$, $\alpha$, and $\Phi$. For this comparison we use as an arbitrary example the results for the meteorite Gilgoin Station. 

In the upper panel the optimum value for the porosity $\Phi$ is held fixed and $s$ and $\alpha$ are varied each in some interval around the optimum values of $s$ and $\alpha$. In the middle panel the optimum value for the aspect ratio $\alpha$ is held fixed and $s$ and $\Phi$ are varied around the optimum values of $s$ and $\Phi$. In the bottom panel the optimum value for $s$ is held fixed and $\Phi$ and $\alpha$ are varied around the optimum values of $\Phi$ and $\alpha$. Also drawn are the contour lines for $\chi^2/\chi^2_\mathrm{min}=1.05$, 1.5, 5, and 10 and their projections onto the basis plane. For Gilgoin Station the minimum value taken by $\bar{\chi}^2$ is 0.15 (see Table \ref{TapFitMet}). The contour line for $\chi^2/\chi^2_\mathrm{min}=5$ therefore encloses the region where the condition $\bar{\chi}^2\lesssim1$ is satisfied which is the minimum requirement for that a fit can be considered as acceptable.

\begin{figure*}

\includegraphics[width=.48\hsize]{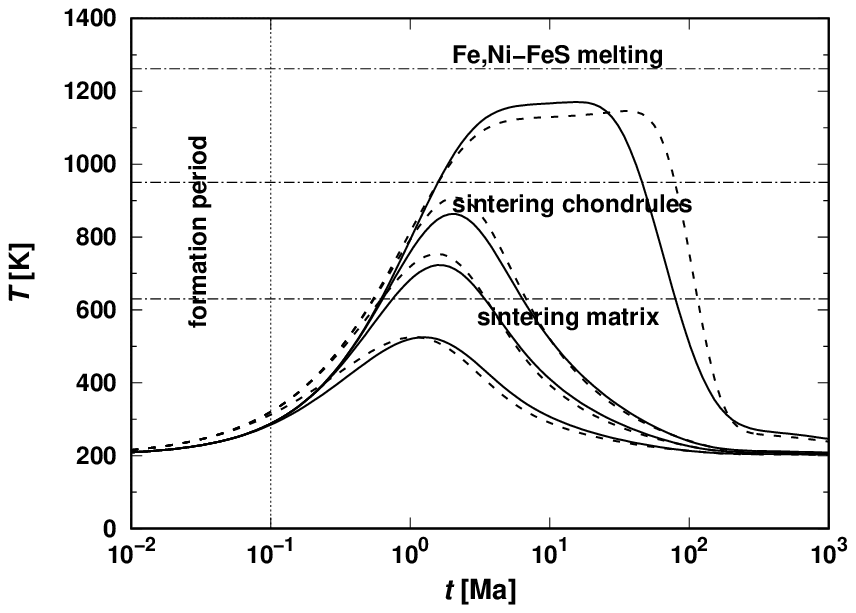}
\includegraphics[width=.48\hsize]{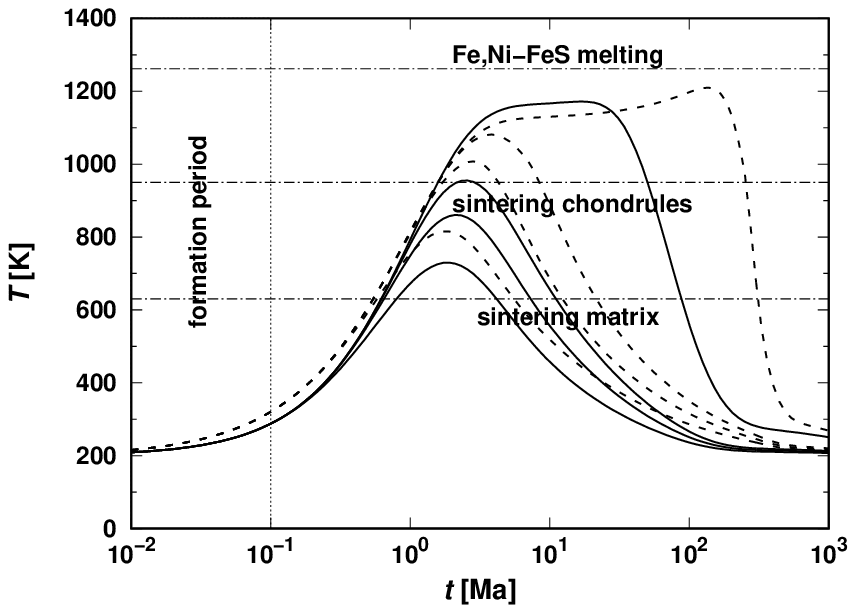}

\caption{Models for the temperature evolution of an asteroid with 150 km radius at different depth below the surface (from bottom to top: 5 km, 10 km, 15 km, and at centre) using different assumptions with respect to the heat conductivity. 
\emph{Left panel:} Homogeneous body. Solid line: Constant heat conductivity $K=4.9$ W\,m$^{-1}$K$^{-1}$ and constant heat capacity $c_p=865$ J\,g$^{-1}$. Dashed line: Variation of heat conductivity with temperature $T$ as $K=4.9\cdot(300\,\mathrm{K}/T)^{1/2}$ W\,m$^{-1}$K$^{-1}$ as typical for rock material \citep{Xu04}, and variation of heat capacity as $c_p=800+0.25\cdot T-1.5\,10^7/T^2$ typical for H-chondrite material \citep{Yom84}.
\emph{Right panel:} Heterogeneous structure with compact core and a 2 km thick insulating mantle. Solid line: Constant heat conductivity $K=4.9$ W\,m$^{-1}$K$^{-1}$ and constant heat capacity $c_p=865$ J\,g$^{-1}$. Mantle with $K$ reduced by a factor of 5, Dashed line: Heat conductivity for H-chondrite material according to the heat conductivity model of this work and variation of heat capacity with $T$ as for H-chondrite material. The time axis corresponds to age after formation. The body ist assumed to be formed at 2.2 Ma after CAI formation. The composition corresponds to that of H chondrites.
}

\label{FigCompMod}
\end{figure*}

An inspection of Fig.~\ref{FigDepPar} shows that the value of the porosity $\Phi$ is rather well restricted by the optimisation process. Rather small deviations from the optimum value result already in considerable changes of $\bar{\chi}^2$. The values of $\alpha$ and $s$, however, can be varied over a considerable range before one observes strong variations of $\bar{\chi}^2$. The anticorrelation in the dependence of $\bar{\chi}^2$ on $\alpha$ and $s$ results from the fact that an increase of $\alpha$ or $s$ both results in an increase of the heat conductivity such that an increase in one of them requires a reduction of the other one in order not to worsen the fit. 

In view of this we tried a fit of the data by using a fixed value of $\alpha=25$ which is roughly the average value of the optimum fit values for all meteorites. Except for a few cases the resulting value of $\bar{\chi}^2$ obtained by varying $\Phi$ and $s$ and $\alpha$ held fixed changes by a few percent only, i.e., the quality of the fit that can be obtained depends only weakly on the precise value of $\alpha$ used for the fit as long as it is within the range of typical values. The value of $s$, however, changes in that case considerably to compensate for the deviation of $\alpha$ from its value taken in the true optimal fit. 

An analogue behaviour is not observed, however, if one fixes $s$ to a value of, for instance, $s=70$, corresponding to the average of the values taken by the optimum fit (omitting the three cases where $s>1\,000$). In nearly all cases one observes strong increases of the new value $\bar{\chi}^2$, obtained if one fits the observed values of $K(T)$ by our model using variable $\Phi$ and $\alpha$ and fixed $s$, over its optimum value obtained by fitting with varying all three parameters. The reason for this behaviour is the fact, that reducing the wall scattering length changes the shape of the heat conductivity curve while a variation of $\alpha$ results in a temperature-independent change. The shape of the curve can only be reproduced by an appropriate choice of the value of $s$.  

The strong deviation of the heat conductivity of chondrites from its expected value for a compact material and the strong differences observed even for meteorites from the same class and petrologic type are most sensitively determined by both, the porosity (given by the porosity of the granular material plus the impact-induced porosity in form of micro-cracks) and the reduction of phonon scattering length by micro-cracks. Variations of the shape of the pores have only a minor influence on the effective heat conductivity of the chondritic material; their influence can easily be mimicked by adjusting $s$. 

\subsubsection{Dependence on depolarisation coefficients}

In all the above fits we have used a unified value of $\alpha$ for each component of the mixture of materials. This needs not be true and probably there are differences to be expected between the different components. This holds in particular for the void space and the iron. The heat conductivities of both are significantly different from that of the mineralic component. 

The minerals in the mixture have similar bulk heat conductivities, in particular the iron-poor olivine and orthopyroxene and the clinopyroxene (Table \ref{TabMinProp}). Since they dominate in the mixture the effective heat conductivity will be not far from their heat conductivities. To zero order, their depolarisation coefficients $L$ would cancel from Eq.~(\ref{EffMediumBrugEllip}) and the corresponding terms in the equation would be essentially the same as if we had chosen $L_j=\frac13$ (i.e., for the case of spherical inclusions) for all of them. 

Such a cancellation effect cannot be expected, however, for the pore space and the iron, because oblate pores act as strong obstacles to heat flow if the normal vector to their flat side is parallel to the local temperature gradient, while iron inclusions act because of their high heat conductivities as short circuit if the normal vector is orthogonal to the temperature gradient. We also made a set of fits where only the depolarisation coefficients of the voids and of iron are calculated according to Eq.~(\ref{Depolar}) for flat oblate rotational ellipsoids with aspect ratio $\alpha$ which is varied, while the depolarisation coefficients of the minerals are set to that of spherical inclusions. As to be expected, the results are almost the same as in the case where depolarisation coefficients for oblate rotational ellipsoids are used for all components, because the depolarisation coefficients of the silicate components nearly cancel. 

If, however, the depolarisation coefficients for the voids or for the iron inclusions are set to one third, the resulting fits become unreliable ($\bar{\chi}^2\gg1$). It is important for explaining the observed heat conductivities that the shape of the micro-cracks and of the iron inclusion is strongly non-spherical. In principle one even could (our should) consider different values of $\alpha$ for the voids and for the iron inclusions, but in our opinion this would overstress the simple model that we use here.   
 
\section{Discussion}

The results of heat conductivity measurements for chondritic material are generally used for estimating the heat conductivity of asteroid material and for calculating on this basis the temperature field in asteroids or fragments of them, or for calculating their thermal evolution. Comparing model calculations using either constant heat conductivity throughout the whole body or considering an additional outer insulating layer of lower heat conductivity \citep[see, e.g.,][ for typical results]{Miy81,Akr97,Tri03,Hen11} one readily sees that the internal thermal structure of asteroids is very different in both cases.  With constant heat conductivity one observes a rather uniform decrease of temperature between the centre of the body and its surface, while the existence of an outer insulating layer results in a slow decline of temperature from the centre to the lower boundary of the insulating layer and a rapid temperature drop to the cold surface within the insulating mantle. Therefore, all conclusions with respect to, e.g., differentiation and mineral composition of the body crucially depend on the kind of variation of heat conductivity within the body.

As example Fig.~\ref{FigCompMod} shows temperature histories at three depths for surface-near layers and at the centre of the body for models with and without insulating mantle. The properties of the body are chosen such that it corresponds to the putative parent body of the H chondrites \citep[cf.][]{Hen16}. By inspecting the figure one readily recognizes how strongly the temperatures in particular in the surface-near layers are increased by the presence of the insulating layer and how important it is to include this in model calculations.

An insulating surface layer may have its origin in two kinds of processes:

First, during the initial build-up of the body from the dust material of the pristine accretion disk by whatsoever process a rather porous body is expected to form during a short growth period of no longer duration than a few 10$^5$ years, which over a more extended period of the order of about $10^6$ years is heated by decay of $^{26}$Al and some other short- and long-lived radionuclides. This results in rapid compaction by sintering and re-crystallisation of the porous initial mixture of dust or dust and chondrules \citep{Yom84,Hen11} once the temperature exceeds a (material dependent) critical temperature for sintering. If the asteroid forms early and $^{26}$Al heating is substantial, this happens in less than about 3 Ma. After compaction the heat conductivity equals that of rock material and is significantly higher than the porous precursor material which has heat conductivities lower by factors up to three compared to rock material and equals more that of, e.g., sand or sandstone \citep[as discussed in detail in, e.g.,][]{Hen16}. The final result is a structure with a rocky core and a sandstone- or sand-like mantle of ca. one to a few km thickness \citep{Yom84,Akr97,Hen11}.

Second, after Jupiter formation the r.m.s velocity of planetesimals in the asteroid belt is stirred up within about 2 Myr to values of 5  \dots\ 10 km\,s$^{-1}$  and the surface structure of the bodies becomes heavily modified by strong impacts and cratering \citep[see, e.g.,][]{Wal11,Dav13,Joh16}. This has two consequences: At one hand, a regolith layer of possibly some km thickness develops at the surface. On the other hand, strong shocks from high-velocity impacts running into the body leave behind numerous micro-cracks \citep[e.g.][]{Ahr02}. 

The possible impact of a regolith layer on thermal evolution was studied by \citet{Akr97} and \citet{Har10} on the basis of the analytic fit of \citet{Yom84} for the $\Phi$ and $T$ dependence of the heat conductivity of chondritic material, which predicts a similar reduction of the heat conductivity of regolith compared to rock material as for sand or sandstone. The time dependent build-up of a regolith layer on bodies in the asteroid belt by impacts was studied by \citet{Hou79} and by \citet{War02}. Unfortunately, both are based on outdated low impact rates and the results cannot be used to estimate how rapid and to what a thickness a regolith layer develops over the about first 100 Ma which determine the basic structure and composition of the asteroids and protoplanets. Since rather thick regolith layers could be built-up, such a layer may be more efficient in blocking the heat flux from the interior of the body to its surface than the residual porous layer that escapes the sintering process. An updated study for the time dependent evolution of a regolith layer is required. The importance of thermal fatique and impacts of micrometeorites as a process for the formation of cracks and of fine material on the surface of small bodies was discussed by \citet{Del14} and \citep{Bas15}.

In any case, micro-cracks are much more efficient for reducing the heat conductivity of rocky material. This was found in the study of \citet{Hen16} where the impacts of macro-porosity and of micro-cracks on heat conductivity were studied. In the present study we find in line with this also that the heat conductivity of all meteorites for which such laboratory measurements are available are mainly ruled by crack-geometry and crack-density. 

The crucial role that micro-cracks play for the heat conductivity, $K$, of meteoritic material means that the value of $K$ measured for a particular meteorite mainly reflects the local impact history at the special location at the surface of an asteroid, or of some fragment of it, where this meteorite was excavated from, and not an intrinsic property of the material from which the corresponding parent body is built. For the limited number of meteorites for which such data are presently available the observed values show no clear dependence on the meteorite type, despite significant compositional differences for different meteorite classes. This makes it difficult to extract from the experimental data general relations for the dependency of the heat conductivity on parameters. Nonetheless, some general relations describing the dependency of $K$ on porosity, $\Phi$, and temperature, $T$, have been given by \citet{Yom84} and for a wide range of porosities by \citet{Hen11}, which were used in a number of model calculations for the thermal evolution of asteroids \citep[e.g.][]{Ben96,Akr97,Hev06, Har10,Hen11, Hen12, Hen13,Gol14,Lic16}. Fits to their experimental results for the temperature variation of $K$ have been given by \citet{Ope12}.

The eventualities connected with the impact history even makes it unlikely that such general relations can exist. The heat conductivity should at least also depend on the extent to which the material properties are modified by shocks, which is usually quantified by the shock stage S1, \dots, S6 \citep{Sto91}. The number of meteorites with measured heat conductivity for which a shock classification could be found in the literature is small, however. The few samples are listed in Table~\ref{TapFitMet}. These are far too few to detect any systematics. Moreover, it is not expected that there exists any clear correspondence between shock stage and density of micro-cracks, because cracks are mainly the result of more frequently occurring weak shocks (i.e., a few GPa maximum pressure), while the shock stage records only the strongest shock event (e.g., tens of GPa). Finally, shock features may be annealed by high-temperature excursions due to very strong shocks.

Our investigation shows that the observed variability of the heat conductivity is not so much due to the chemical/mineralogical differences between different meteorite classes but is essentially ruled by the strongly enhanced phonon-scattering at micro-cracks, which can be accounted for by a variation of only one parameter (our parameter $s$). The variation of heat conductivity of meteorites, therefore, probably would be best described by a frequency distribution for the value of the parameter $s$. Then it would be possible to derive a sound average value for the heat conductivity in that layers of a meteorite parent body where matter is strongly impact-modified, which then can be used in model calculations.

Presently too few data are available to construct an empirical frequency distribution. It would certainly be possible to derive such a distribution theoretically if the relation between shock strength and density of micro-cracks produced is known. Then combining studies of the evolution of impact-cratering of asteroid surfaces over time \citep[like, e.g., that of][]{Dav13} with theoretical models of impact induced shocks \citep[e.g.][]{Dav12} would allow to construct the required distribution. But presently no such studies are available.

We have not considered one aspect of the problem: The anisotropy of the heat conductivity. In the experiments of \citet{Ope12} the heat conductivity of the meteorites Bath Furnace and Holbrook have been obtained for different directions of heat flow with respect to the orientation of the sample. Considerable differences were found for different directions. This directional dependence of heat conductivity observed for the meteorites is probably due to aligned orientation of the mineral crystallites in the polycrystalline material and to aligned orientation of the micro-cracks, as a result of shocks \citep{Gat05}. The present stage of modelling the structure and evolution of asteroids presently does not allow to include such effects, so we leave aside such complications.


\section{Concluding remarks}
\label{SectConclu}

We computed the temperature dependent heat conductivity of the granular mixture of minerals and tiny lumps of a NiFe-alloy that is found in ordinary chondrites. For this purpose we fitted to the model of \citet{Cal59} for the temperature dependence of the conductivity of crystalline materials laboratory data for the heat conductivity of pure crystalline substances collected from the literature. From this we calculated the heat conductivity of  the different mineral components in chondrites over a wide temperature range. We, then, applied the Bruggeman mixing rule \citep[see, e.g., ][]{Boh83} for multi-component mixtures to calculate the effective heat conductivity of the mixture of minerals, metal-alloy, and voids found in ordinary chondrites. 

We compared this model for the temperature variation of the heat conductivity with published results of laboratory measurements of the temperature variation of H, L, and E chondrites. A considerable discrepancy between model and experimental results was found, if material properties of unshocked materials were used in the heat-conductivity model. If the damage of the crystalline structure in form of the numerous micro-cracks generated by strong shocks was considered, good agreement between model and experimental results could be obtained, however. 

We conclude from this, that the value of the heat conductivity and the kind of its temperature variation observed for meteorites is largely ruled by the reduction of the mean free path for phonon-scattering in crystals by abundant micro-cracks and by the large aspect-ratio of such fissures in the solids.

Our model provides a natural explanation for the observed high variability of heat conductivities of meteorites with otherwise largely identical properties. These variations reflect the varying degree by which their lattice structure is damaged by shocks, introducing numerous internal walls for phonon-scattering.

This means that laboratory measured data for heat conductivities of meteorites cannot easily be taken as guides for estimating the heat conductivity of asteroid material for simulations of their thermal structure and evolution. They reflect a local surface property at the place where a meteorite was excavated and depend heavily on the eventualities of the impact history at that location. They do not provide information on the properties of the material in deeper layers which are not strongly affected by shocks. Heat conductivity of rock material would be a much better guideline at higher depths, which has not always been observed in past model calculations. 

Our present model allows to calculate heat conductivities for both, shocked and unshocked chondritic materials and can also be applied to not-yet laboratory studied cases and, more generally, to arbitrary meteorite compositions that might exist elsewhere. 

The model has, however, also some shortcomings. The temperature dependence of the heat conductivity of the main minerals in chondrites seemingly has never been studied at very low temperatures. At temperatures much lower than 300 K our model may be of low accuracy, since due of the lack of low-temperature data it relies on an approximation of the Debye-temperature for heat conductivity. On the other hand, compared with the data of \citet{Ope10,Ope12} at temperatures below 100 K, the deviation between the model results and measured data seems to be insignificant, i.e., the model seems to work well even in that region where it presently is not sufficiently underpinned by experimental data. The only exception are the enstatite chondrites. Low-temperature data at least for the enstatite mineral are required for bringing the model closer to reality also for enstatite-dominated chondritic matter.


\begin{acknowledgements}
We greatly acknowledge that G. Consolmagno made available to us the detailed data set for heat conductivity of meteorites from \citet*{Ope12}. This work was performed as part of a project of `Schwerpunktprogramm 1385', supported by the `Deutsche Forschungs\-gemeinschaft (DFG)'.  MT acknowledges support by the Klaus Tschira Stiftung gGmbH. This research has made use of NASA's Astrophysics Data System.
\end{acknowledgements}


\def\mps{Meteoritics~\&~Plan.~Sci.}%
\def\epsl{Earth~\&~Plan.~Sci.~Lett.}%
\def\lpscl{Lunar Planet. Sci. Conf. Lett.}%

\end{document}